\documentclass[pra,twocolumn,aps,showpacs,preprintnumbers,amsmath,amssymb,superscriptaddress]{revtex4-1}

\usepackage{graphicx}
\usepackage{dcolumn}
\usepackage{bm}
\usepackage{amsmath}

\begin{document}


\title{Implementing the Deutsch-Jozsa algorithm with macroscopic ensembles}

\author{Henry Semenenko}
\affiliation{The University of Nottingham, University Park, Nottingham, NG7 2RD, UK}
\affiliation{The University of Bristol, Senate House, Tyndall Avenue, Bristol, BS8 1TH, UK}

\author{Tim Byrnes}
\affiliation{New York University, 1555 Century Ave, Pudong, Shanghai 200122, China}
\affiliation{NYU-ECNU Institute of Physics at NYU Shanghai, 3663 Zhongshan Road North, Shanghai 200062, China}
\affiliation{National Institute of Informatics, 2-1-2 Hitotsubashi, Chiyoda-ku, Tokyo 101-8430, Japan}

\date{\today}

\begin{abstract}
Quantum computing implementations under consideration today typically deal with systems with microscopic degrees of freedom such as photons, ions, cold atoms, and superconducting circuits.  The quantum information is stored typically in low-dimensional Hilbert spaces such as qubits, as quantum effects are strongest in such systems.  It has however been demonstrated that quantum effects can be observed in mesoscopic and macroscopic systems, such as nanomechanical systems and gas ensembles. While few-qubit quantum information demonstrations have been performed with such macroscopic systems, a quantum algorithm showing exponential speedup over classical algorithms is yet to be shown.  
Here we show that the Deutsch-Jozsa algorithm can be implemented with macroscopic ensembles. The encoding that we use avoids the detrimental effects of decoherence that normally plagues macroscopic implementations. We discuss two mapping procedures which can be chosen depending upon the constraints of the oracle and the experiment.  Both methods have an exponential speedup over the classical case, and only require control of the ensembles at the level of the total spin of the ensembles.  It is shown that both approaches reproduce the qubit Deutsch-Jozsa algorithm, and are robust under decoherence.  
\end{abstract}

\pacs{03.75.Gg, 03.75.Mn, 42.50.Gy, 03.67.Hk}
\maketitle

\section{Introduction}

In 1985, David Deutsch provided the first quantum algorithm that showed the potential for quantum computing to be more powerful than classical computing -- Deutsch's algorithm \cite{deutsch1985}. It was later expanded upon by Richard Jozsa \cite{deutsch1992}, and improved by Cleve, Ekert, Macchiavello and Masca \cite{cleve1998}, resulting in the multi-qubit generalization known today as Deutsch-Jozsa algorithm. One of the key features of the algorithm is that creates a superposition of all possible states, then is followed by an interference and measurement step -- a key component of many quantum algorithms.  It therefore paved the way for other more practical quantum algorithms such as Grover's algorithm \cite{grover1996} and Shor's algorithm \cite{shor1994}, both of which provide a quantum mechanical speedup over the best available classical algorithms. For this reason the algorithm remains important from a theoretical point of view of the power of quantum computing, and from an experimental point of view as a proof-of-principle operation of quantum computer prototypes.  Examples of experimental demonstration of Deutsch-Jozsa algorithm include NMR \cite{collins2000,wu2011}, superconducting qubits \cite{schuch2002}, single photon linear optics \cite{takeuchi2000}, and trapped ions \cite{gulde2003}.  Currently demonstrations of quantum algorithms are typically limited to $ \lesssim 10 $ qubits, due to limitations with decoherence and scalability of current quantum computing technologies. 

To implement a given quantum algorithm, currently there are two main paradigms of quantum computation, using either discrete or continuous variables (CV).  The most commonly used approach uses discrete quantum states to encode quantum information, typically in the form of qubits.  Alternatively, one may store quantum information in a bosonic mode which has an infinite Hilbert space dimension, and states can be visualized in the phase space of position and momentum \cite{braunstein05}.  An equivalent approach involves using total spin operators as quasi-bosonic variables to implement CV \cite{julsgaard01,krauter12}. Recently, a third alternative to these paradigms has emerged \cite{byrnes2012,byrnes2014}, having characteristics common to both.  The scheme -- which we call ensemble quantum computation (EQC) -- stores quantum information on ensembles of qubits and manipulates them using only products of total spin operators.  While it has been known for some time that it is possible to form continuous variable bosonic mode operators using polarized spins, the scheme differs from this by the full use of the space of states available on the Bloch sphere.  For continuous variables implementations typically the spins are polarized in the $ S^X $ direction and only small deviations from this are induced.  The scheme has the advantage that it has the same Bloch sphere structure as is the case with standard qubits, yet with a natural robustness due to the use of ensembles instead of single qubits.  As the states that are used are not explicitly gaussian in a CV sense, many of the no-go results for continuous variables do not immediately apply, making non-trivial operations possible with low order products of spin operators.  Indeed, it has been shown that universal operations are possible with products of one and two total spin operators \cite{byrnes2014}.  

One of the difficulties with EQC is that it is not always straightforward to translate a qubit or CV quantum algorithm into that with ensembles. Part of the difficulty here is that due to the large Hilbert space available to the ensemble as compared to the original qubit circuit, the mapping is not unique.  Thus there is a great amount of freedom in choosing the best encoding of the original problem in the ensemble case, and the best way to do this.  By ``best'' way, this includes considerations such as: (i) requiring no complicated Hamiltonians beyond low-order products of total spin operators; (ii) the output of the quantum algorithm is not adversely affected by the generation of unstable quantum states such as Schrodinger cat states; (iii) the performance of the algorithm (as measured by e.g. success probability, fidelity, etc.) remains the same or acceptably high under realistic conditions.  For these reasons, the mapping between qubit algorithms to EQC requires some analysis, and currently no general procedure exists to map between the two.  Nevertheless, to date several algorithms have been shown to be mapped, incuding quantum teleportation \cite{pyrkov14,pyrkov14b} and Deutsch's algorithm \cite{byrnes2014}.  

In this paper we provide a full analysis of mapping the Deutsch-Jozsa algorithm to EQC. Our aim is to start with the qubit version of the algorithm, and convert this to an implementation using ensembles and ultimately a Hamiltonian involving only products of total spin operators.  As mentioned above, as the EQC mapping involves mapping qubits onto ensembles, there are in fact many possible mappings which in principle accomplishes the task.  Partly to this reason, we find two viable mappings, which are both presented in this paper. This paper is structured as follows. In Sec. \ref{sec:ensembleqc} we give a review of the EQC framework, in the interest of this paper being self-contained.  We then review the Deutsch-Jozsa algorithm for qubits in Sec. \ref{sec:djforqubits}, which serves to introduce our notation.  Due to the rather detailed nature of this paper, we then summarize in Sec. \ref{sec:djforeqc} our final results for how to map the Deutsch-Jozsa algorithm onto EQC for readers who are not interested in the details of the proof. The remaining sections are devoted to the proof of how the EQC mapping works.  One of the results that we will require is an explicit form of the oracle Hamiltonian for qubits, which is derived  in Sec. \ref{sec:oracle}. We show that some of these implementations are more favorable for the EQC than others.  In the case that it is possible to choose exactly how the oracle is implemented, a mapping that is robust against decoherence for EQC is presented (Sec. \ref{sec:method2}).  In the case that the oracle is strictly not choosable, and it must be mapped directly from the qubit case, we provide another mapping which works for all cases (Sec. \ref{sec:method1}).  We finally summarize our findings in Sec. \ref{sec:conc}.

\section{Ensemble quantum computation}
\label{sec:ensembleqc}

In this section we provide a brief summary of the essential aspects of EQC, for the benefit of this paper being self-contained.  A more detailed description is given in Ref. \cite{byrnes2014}.  

In EQC, quantum information is stored on ensembles of two level systems.  This can be either a large number of individual qubits such as an atomic ensemble, or a two-component Bose-Einstein condensate (BEC) \cite{bohi09,riedel10,byrnes2012,byrnes2014}.  In this approach, the quantum information corresponding to a qubit $ |\psi \rangle = \alpha | 0 \rangle + \beta | 1 \rangle $ with $ | \alpha |^2 + | \beta |^2 = 1  $ is stored as a spin coherent state.  For an atomic ensemble, this is written
\begin{align}
 |\alpha,\beta\rangle\rangle \equiv \prod_{m=1}^N (\alpha | 0 \rangle_m + \beta | 1 \rangle_m )
 \label{eq:ensemblestate}
\end{align}
where $ | 0\rangle_m $ and $ | 1 \rangle_m $ are the logical states of the $ m $th qubit in the ensemble. In the case of a 
BEC, the spin coherent state is
\begin{align}
 |\alpha,\beta\rangle\rangle \equiv \frac{1}{\sqrt{N!}}(\alpha a^\dagger + \beta b^\dagger)^N|0\rangle
 \label{eq:BECstate}
\end{align}
where $ a, b $ are bosonic annihilation operators satisfying $[a,a^\dagger]=[b,b^\dagger]=1$ corresponding to the two logical states that store the quantum information.  In each case we assume a fixed number of particles $ N $ in the ensemble or BEC.  

The states in (\ref{eq:ensemblestate}) and (\ref{eq:BECstate}) may be expanded in terms of Fock states with definite particle number.  For the ensemble system we may define
\begin{align}
| k \rangle = \frac{1}{\sqrt{N \choose k }} \sum_{ \underset{ \{ \sum_m x_m = N-k \} }{x_1 x_2 \dots x_N }} | x_1 x_2 \dots x_N \rangle ,
\label{fockstatesens}
\end{align}
where $ x_m \in \{0,1 \} $ and the sum is restricted states with $ N-k $ spins in the state $ | 1 \rangle $ and $ k $ in the state $ | 0 \rangle $.  For the BEC case, the Fock states are
\begin{align}
| k \rangle = \frac{1}{\sqrt{k! (N-k)!}} (a^\dagger)^k (b^\dagger)^{N-k} | 0 \rangle  .
\label{fockstates}
\end{align}
The spin coherent states (\ref{eq:ensemblestate}) and (\ref{eq:BECstate}) can be expanded using Fock states into 
\begin{align}
 |\alpha,\beta\rangle\rangle = \sum_{k=0}^N \sqrt{N \choose k } \alpha^k \beta^{N-k} | k \rangle,
\end{align}
which is true for both the ensemble and BEC cases.  

For manipulation of the state (\ref{eq:ensemblestate}) and (\ref{eq:BECstate}) we use the total spin operators 
\begin{align}
S^X & = \sum_{m=1}^N \sigma^X_m  \nonumber \\
S^Y & = \sum_{m=1}^N \sigma^Y_m  \nonumber \\
S^Z & = \sum_{m=1}^N \sigma^Z_m 
\label{ensemblespins}
\end{align}
where $ \sigma^{X,Y,Z}_m $ are the Pauli operators for each qubit in the ensemble, defined according to 
\begin{align}
\langle x'  | \sigma^X | x \rangle & = \delta_{x ,1-x'}  \nonumber \\
\langle x'  | \sigma^Y | x \rangle & = i  (-1)^x  \delta_{x, 1-x'}  \nonumber  \\
\langle x'  | \sigma^Z | x \rangle & = (-1)^x \delta_{x, x'}  
\end{align}
where $ x \in \{ 0,1 \} $ and $ \delta_{x, x'} $ is the Kronecker delta. For the BEC case, the total spin operators are Schwinger boson operators
\begin{align}
S^X & = a^\dagger b + b^\dagger a \nonumber \\
S^Y & = -ia^\dagger b + ib^\dagger a  \nonumber \\
S^Z & = a^\dagger a - b^\dagger b 
\label{bosonicspins}
\end{align}
The Fock states are eigenstates of the $ S^Z $ operator, both for the ensemble and BEC cases we have
\begin{align}
S^Z | k \rangle = (2k - N ) | k \rangle .
\end{align}

The total spin operators obey the same commutation relations as Pauli operators
\begin{align}
[ S^i, S^j ] = 2 i \epsilon_{ijk} S^k
\label{commutation}
\end{align}
where $ \epsilon_{ijk} $ is the Levi-Civita antisymmetric tensor.  While (\ref{commutation}) suggests an analogous structure to standard qubits, the total spin operators do not satisfy
\begin{align}
\{ S^i, S^j \} \ne 2 \delta_{ij}
\label{anticommutation}
\end{align}
where $ \delta_{ij} $ is the Kronecker delta.  For qubits $ N= 1 $, the anticommutation relation is satisfied, which in many cases results in simplifications.  For example, (\ref{anticommutation}) implies that $ (\sigma^{X,Y,Z})^2 = 1 $, which is not true for the $ N \ge 2 $ case.  For our calculations we will generally use the BEC formulation of the total spins (\ref{bosonicspins}) rather than the ensemble formulation (\ref{ensemblespins}) for the sake of mathematical simplicity.  In fact these are equivalent as long as all physical operations and the initial conditions of the spins are symmetric under particle interchange.  Thus either ensembles or BECs could be used experimentally. 

The aim of EQC is then exploit the analogous structure of the spin coherent states to qubits to provide a framework for quantum computation. In the same way as qubits and CV approaches where many qubits and modes are used to store the quantum information, in a typical EQC algorithm one would use several ensembles, which are potentially entangled together.  A typical entangling interaction between ensembles that is considered is a $ H = S^Z_1 S^Z_2 $ interaction, as this can be implemented experimentally using several schemes \cite{pyrkov2013,hussain2014,abdelrahman14}.  Such an interaction produces in general a complex entangled state, exhibiting entanglement with a fractal structure \cite{byrnes13,kurkjian13}.  Nevertheless for particular gate times this has a simplified structure which may be used for quantum information tasks \cite{pyrkov14,pyrkov14b}.  

One of the advantages of EQC is that quantum information is always stored in a highly duplicated way. This allows for a more robust storage of quantum information as the loss or corruption of a few of the particles making up the ensemble impacts the total spin in a negligible way.  This is in contrast to single particle storage methods where one qubit's worth of information is stored on one physical qubit.  In this case if the particle is lost or an error occurs, all the quantum information is destroyed, which motivates quantum error correction.  Another benefit is that experimentally manipulating ensembles is an easier task technically compared to single particles, with the additional benefit of an increased signal to noise in any measurement readout.  Although not used in this paper, another benefit is that ensembles have the possibility of non-destructive readout, an operation which is fundamentally not possible with qubit based systems \cite{byrnes14,ilookeke2014}.

Due to the larger Hilbert space available to the ensembles, given a quantum algorithm intended for qubits, in principle there are many ways to map it only the ensemble system.  For example, one simple way would be to pick two states in the ensemble and use this as the logical states.  However, such an approach would not be desirable as it would be experimentally challenging to target two particular states in the ensemble. By this we mean that exotic gates with complex Hamiltonians are required. 
 It is also potentially susceptible to decoherence.  For example, using an encoding of states such as $ | 0_L \rangle = | 0 0 \dots 0 \rangle $ and $ | 1_L \rangle = | 1 1 \dots 1 \rangle $ would correspond to using Schrodinger cat-like states, which are vulnerable to decoherence.  

For these reasons we impose the following additional restrictions and assumptions when constructing a quantum algorithm in EQC:
\begin{itemize}
\item Only gates involving Hamiltonians with linear products of total spins $ S^{X,Y,Z} $ are used. 
\item Measurements are made in a collective basis, e.g. $ S^Z $. 
\item The performance of the algorithm should not degrade exponentially with particle number $ N $ under decoherence.
\item The gate resource count for applying a gate to an ensemble is the same as for a qubit.
\end{itemize}
The first and second restrictions ensure that any algorithm constructed should be able to be implemented using reasonable means.  As discussed in Ref. \cite{abdelrahman14}, collective operations are typically of the form of linear products of total spin operators $ S^{X,Y,Z} $.  The third restriction requires analysis of the quantum algorithm under the presence of decoherence.  For example, if the algorithm generates Schrodinger cat states and stores quantum information that affects the quantum algorithm, this could adversely affect the performance.  

The last assumption is important from the point of view of whether a quantum algorithm has been mapped correctly with a quantum speedup.  Since an ensemble involves $ N $ individual qubits, a question arises of whether we count resources on a per qubit or per ensemble basis. In our method, we take the latter approach for the reason that only collective operations are performed on the ensembles.  When applying a collective operation to ensembles, we assume that it is no more difficult (i.e. experimentally time consuming) to perform the operation on the ensemble as compared to the qubit.  For example, in the case of the optical manipulation of Ref. \cite{abdelrahman14}, a Raman laser pulse performs an $ S^X $ rotation of the ensemble.  In this case, the time required in order to rotate one qubit compared to $ N $ qubits is the same, as the same laser pulse illuminates all atoms simultaneously.  Counted in this way, we consider a single gate to operate on all the qubits within an ensemble in parallel, such that gate resource counts are the same for an ensemble and a single qubit.

\section{Deutsch-Josza algorithm}

\subsection{Qubit implementation} 
\label{sec:djforqubits}

In this section we review the Deutsch-Josza algorithm for qubits, which will serve to introduce our notation and highlight several aspects of the algorithm which will be useful later.  In particular, we derive explicit expressions for the Hamiltonian of the oracle, which plays a central role in the algorithm.  

Consider a function $ f(x) $ which takes an integer input $ x \in [0,2^M-1] $ and outputs a binary result $ f \in [0,1] $ (see Fig. \ref{fig1}a). The types of functions that are allowable to two types.  The first type, called ``constant'', has an output which is constant for all $ x $.  There are only two types of constant functions, $ f = 0$ and $ f = 1 $.  The second type, called ``balanced'', has exactly half its output being 0 and the other half being 1. There are $ {2^M \choose 2^{M-1}} $ such balanced functions.   Now consider that we are given a device, the ``oracle'', that implements the function $ f(x) $ according to the above restrictions.  The aim of the Deutsch-Jozsa algorithm is to discriminate a given function $ f(x) $ between the balanced and constant cases, with as few calls to the oracle as possible. Classically, to make this classification with certainty, it is necessary to call the oracle more than half the number of input values, i.e. $ 2^{M-1} + 1 $ times.  

Quantum mechanically, it is possible to speed this up exponentially.  The oracle is implemented such that it follows a relation
\begin{align}
U_f  | y\rangle | x \rangle =  | y \oplus f(x) \rangle | x \rangle
\label{oracleunitary}
\end{align}
where $ y \in [0,1] $ and $ \oplus $ is the logical XOR gate.  The $ x $-register consists of $ M $ qubits in a binary representation, as shown in Fig. \ref{fig1}b. Assuming that the oracle can take a superposition of input states, then the quantum circuit Fig. \ref{fig1}b achieves the objective with only one call of the oracle \cite{nielsen00}. For the case that $ f(x) $ is constant, the measurement yields a result with all the $ x $-register qubits in the state $ | 0 \rangle $.  For $ f(x) $ balanced, the measurement yields a result with at least one of the $ x$-register qubits in the state $ | 1 \rangle $.  We summarize the effect of the circuit Fig. \ref{fig1}b as
\begin{align}
|1 \rangle | x= 0 \rangle  \rightarrow \left\{
\begin{array}{ll}
\left( \frac{| 0 \rangle - |1 \rangle}{\sqrt{2}} \right) | x= 0 \rangle  & f(x) \in \mbox{ constant} \\
\left( \frac{| 0 \rangle - |1 \rangle}{\sqrt{2}} \right)  | x> 0 \rangle  & f(x) \in  \mbox{ balanced}  
\end{array}
\right. 
\end{align}
The output of the $ x $-register unambiguously discriminates between the constant and balanced cases, which achieves the objective of the Deutsch-Jozsa algorithm. 

In this paper, we distinguish between two modes of oracle operation: ``classical'' and ``quantum''.  When the oracle has no superposition states as its input, such as in (\ref{oracleunitary}), we call the oracle to be operating in ``classical'' mode.  When the inputs are in a superposition state, such as that shown in Fig. \ref{fig1}b, we say that the oracle is operating in ``quantum'' mode.  There is no difference to the operation of the oracle itself in either case -- the only difference is what states are input to the oracle.

\subsection{EQC implementation}
\label{sec:djforeqc}

\begin{figure}
\includegraphics[width=\columnwidth]{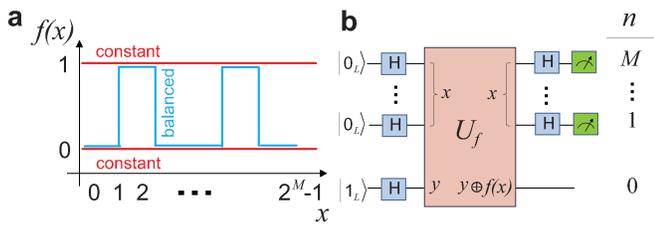}
\caption{The Deutsch-Jozsa algorithm.  (a) The function $ f(x) $ which determines the oracle.  (b) The quantum circuit for the Deutsch-Jozsa algorithm.  The gates marked by $ H $ are Hadamard gates, $ U_f $ is the oracle, and the meter symbols denote a measurement in $ S^Z $-basis. The labeling of the qubits/ensembles for $ n \in [0,M ] $ is shown. }
		\label{fig1}
\end{figure}

Here we present a summary of the Deutsch-Jozsa algorithm in the EQC framework, for the benefit of readers who are not interested in following the details of the proof in the following sections.  We show two methods of mapping the Deutsch-Jozsa algorithm, which are distinct in the way the oracle and quantum information are defined.  In both cases, the basic procedure is to follow the quantum circuit of Fig. \ref{fig1}(b).  The definition of each of the components in the circuit are however different, and are defined in Table \ref{tab1}.  

We point out that as seen in Table \ref{tab1}, all the gates performed are collective operations, involving a Hamiltonian of only linear powers of total spin operators $ S^{X,Y,Z} $. The Hadamard gate for EQC reads as
\begin{align}
a & \rightarrow \frac{a-b}{\sqrt{2}} \nonumber \\
b & \rightarrow \frac{a+b}{\sqrt{2}} ,
\end{align}
which corresponds to applying a Hamiltonian $ S^Y $ for a time $ \pi/4 $.  Similarly, the measurement is in the collective
basis of the eigenstates of the $ S^Z $ operators.  This is one of the requirements of EQC, such that it can be realistically implemented experimentally.

\begin{table*}[t]
\begin{center}
\begin{tabular}{|c|c|c|}
\hline
Component in Deutsch-Jozsa algorithm & Method 1  & Method 2  \\
\hline
$| 0 \rangle_L $ & $ | 0,1 \rangle \rangle $ &  $ | 0,1 \rangle \rangle $  \\
$| 1 \rangle_L $ & Fock state $ |k_0 \rangle $, with $ k_0 \in \text{odd} $ &  $ | 1,0 \rangle \rangle $  \\
Hadamard Hamiltonian & $ S^Y $ &  $ S^Y $ \\
Oracle Hamiltonian mapping & $ \sigma^Z \rightarrow S^Z + N + 1$ & $ \sigma^Z \rightarrow S^Z/N $ \\
Measurement & $ S^Z $ basis &  $ S^Z $ basis \\
Constant outcome & $  \forall n: | 0,1 \rangle \rangle_n  $ & $ \forall n:  | 0,1 \rangle \rangle_n $  \\
Balanced outcome & Any state orthogonal to $  \forall n: | 0,1 \rangle \rangle_n  $ & Any state orthogonal to $  \forall n: | 0,1 \rangle \rangle_n  $ \\
\hline
\end{tabular}
\caption{Summary of mapping of the Deutsch-Jozsa algorithm for ensemble quantum computation. In both Methods 1 and 2, the quantum circuit of Fig. \ref{fig1}(b) is followed, with the definitions as given in the table. The outcome of the measurement for balanced oracle cases are the converse of the constant cases.   }
\label{tab1}
\end{center}
\end{table*}

\section{The Deutsch-Jozsa Oracle}
\label{sec:oracle}

Typically in the discussion of the Deutsch-Jozsa algorithm the specific implementation of the oracle is left unspecified, as this is the object which we are trying to gain information about. There are in fact an infinite number of ways that the relation (\ref{oracleunitary}) can be performed.  However, for a mapping to EQC it is an important question to understand whether the oracle itself can be implemented using ensembles, as now the input of the circuit Fig. \ref{fig1}b are each ensembles instead of qubits.  What is the meaning of (\ref{oracleunitary}) in an EQC implementation?   To this end, we discuss specifically what qubit Hamiltonian is required for a given $ f(x) $ to realize the oracle.

\subsection{Oracle Hamiltonian}
\label{sec:qubitoracle}

Let us start by first writing an explicit form of the unitary of the oracle according to the definition (\ref{oracleunitary}).  The unitary corresponding to the oracle is  
\begin{equation}
	U_f = \sum^{2^M-1}_{x=0}\left[ (1-f(x))I_0 +f(x)\sigma^X_{0} \right]|x\rangle\langle x| 
	\label{generaloracleunitary}
\end{equation}
Looking at each term in the summation, for any $ x $ with $ f(x) = 0 $, the unitary reduces to 
$ U_f = |x\rangle \langle x| $, which leaves the state of $ y $ unchanged.  For any $ x $ with $ f(x) = 1 $, the unitary is $ U_f = \sigma^X_{0} |x\rangle \langle x| $, which flips the state of $ y $.  We see that for the case of constant $ f(x) $, the results of (\ref{constantzero}) and (\ref{constantone}) are recovered. To make this relation more explicit, let us rewrite (\ref{generaloracleunitary}) according to
\begin{align}
U_f = \sum_{x \notin {\cal F}} | x \rangle \langle x | + \sigma^X_{0} \sum_{x \in {\cal F}} | x \rangle \langle x |
\label{generaloracleunitary2}
\end{align}
where $ {\cal F} $ is the set of all $ x $ that satisfy $ f(x) = 1 $.  

Now let us explicitly write the Hamiltonians which give rise to (\ref{generaloracleunitary2}) first for the constant cases.  In the case that $f = 0 $,  this leaves both $ | x \rangle $ and $ | y \rangle $ unchanged hence this may be implemented by
\begin{align}
U_{f=0} = e^{-i H_{f=0} t} = I
\label{constantzero}
\end{align}
where $ I $ is the identity operator on the whole system.  Assuming throughout for concreteness that the Hamiltonian is always evolved for a time $ t = 1 $, a Hamiltonian that implements this is 
\begin{align}
H_{f=0} = 2 \pi j
\label{fzeroham}
\end{align}
where $ j $ is an arbitrary integer.  

For the function $f  = 1 $, this results in always flipping the $ y $-qubit, which can be implemented by 
\begin{align}
U_{f=1} = e^{-i H_{f=1} t} = \sigma^X_0 .  
\label{constantone}
\end{align}
A Hamiltonian that satisfies this is 
\begin{align}
H_{f=1} = \pi (2j+1)  \left(\frac{\sigma^X_{0} - I_0}{2} \right)
\label{foneham}
\end{align}
again for a time $ t = 1 $ and $ j $ is an arbitrary integer.  Here $ \sigma^X_0, I_0 $ denotes the Pauli and identity operator on the $n=0 $ qubit, following the labels as specified in Fig. \ref{fig1}b. The factor of $ 2j+1 $ reflects the fact that any odd integer may multiply the Hamiltonian with the same effect.  In this case this degree of freedom does not play an important role, however we will see that this gives an important degree of freedom when constructing balanced Hamiltonians. 

Now let us turn to the balanced cases. Before constructing the Hamiltonian for (\ref{generaloracleunitary2}), it is instructive to calculate the Hamiltonian for 
an oracle where only one of the $ x $'s  satisfies $ f(x)= 1$, that is
\begin{align}
f_x (x') = \left\{
\begin{array}{ll}
1 & x'= x \\
0 & x' \ne x
\end{array}
\right. .
\end{align}
In this case the oracle gives
\begin{align}
	U_{x} = \sum_{x' \ne x} | x' \rangle \langle x' | + \sigma^X_{0} | x \rangle \langle x | .
	\label{foneunitary}
\end{align}
Similarly to (\ref{foneham}), we can write 
\begin{align}
H_{x} = \pi (2j_x+1)  \left(\frac{\sigma^X_{0} - I_0}{2} \right) | x \rangle \langle x |, 
\label{hxsimple}
\end{align}
which can be verified to satisfy $ U_x = e^{- i H_x t} $ for a time evolution $ t = 1 $, using 
 $ e^{i A |x \rangle \langle x | } = I + (e^{iA} - I) |x \rangle \langle x | $ for an arbitrary operator $ A $. Here $ j_x $ is an integer that can be independently chosen for each $ x $. Rewriting the projection 
operator in terms of Pauli matrices, this is
\begin{align}
H_{x} & =  \pi (2j_x+1) \left(\frac{\sigma^X_{0} - I_0}{2} \right)  \prod_{n=1}^M \frac{1}{2} \left( 1 + (-1)^{x_n} \sigma^Z_n \right) \nonumber \\
& =  \pi (2j_x+1)  \left(\frac{\sigma^X_{0} - I_0}{2} \right) \frac{1}{2^M} \Big[ 1 + \sum_n (-1)^{x_n} \sigma^Z_n \nonumber \\
& + 
\sum_n  \sum_{n' \ne n}  (-1)^{x_n+ x_{n'}} \sigma^Z_n \sigma^Z_{n'} + \dots + \prod_{n=1}^M (-1)^{x_n}  \sigma^Z_n \Big] \nonumber \\
& =  \pi (2j_x+1)  \left(\frac{\sigma^X_{0} - I_0}{2} \right)  \sum_{z=0}^{2^M-1} \frac{(-1)^{z \cdot x}}{2^M} \prod_{n=1}^M (\sigma^Z_n)^{z_n} ,
\label{expandinghx}
\end{align}
where in the second line we have expanded the product to a sum of various products of Pauli matrices. There are $ 2^M $ terms in the expansion, and each of the terms in the expansion is labeled by $ z \in [0,2^{M}-1] $.  The $ x_n, z_n $ are binary representations of $ x $ and $ z $, where $ n $ is the bit label, and $ x \cdot z = \sum_n  x_n z_n $.  

Once we have the Hamiltonian that gives the unitary (\ref{foneunitary}) for one of the $ x $'s, the Deutsch-Jozsa oracle (\ref{generaloracleunitary2}) can be constructed by multiplying together all the cases that satisfy $ f(x) = 1 $.  That is,
\begin{align}
U_f = \prod_{x \in {\cal F}} U_x .
\end{align}
As all the different $ H_x $ for all $ x $ commute, the total Hamiltonian is simply the sum of those satisfying $ x \in {\cal F} $:
\begin{align}
H_f & = \sum_{x \in {\cal F}} H_x \nonumber \\
& = \pi \left(\frac{\sigma^X_{0} - I_0}{2} \right)  \sum_{x \in {\cal F}} (2 j_x +1 )  | x \rangle \langle x | \nonumber \\
& = \pi \left(\frac{\sigma^X_{0} - I_0}{2} \right)  \sum_{x \in {\cal F}} (2 j_x +1 )  \prod_{n=1}^M \frac{1}{2} \left( 1 + (-1)^{x_n} \sigma^Z_n \right) 
\label{projham}
\end{align}
%
Expanding the product, the Hamiltonian can then be written explicitly as
\begin{align}
H_f =  &   \pi \left(\frac{\sigma^X_{0} - I_0}{2} \right) \sum_{z=0}^{2^M-1} \alpha_z \prod_{n=1}^M (\sigma^Z_n)^{z_n} . 
\end{align}
where
\begin{align}
\alpha_z =  \frac{1}{2^M} \sum_{x \in {\cal F}} (2j_x+1 )  (-1)^{z \cdot x}  .
\label{alphazcoeffs}
\end{align}
It is convenient for later to define the coefficients of the expanded version 
of the Hamiltonian
\begin{align}
H_f & =   \pi \left(\frac{\sigma^X_{0} - I_0}{2} \right) \Big[ \alpha_0
+ \sum_n \alpha_n \sigma^Z_n + \sum_n \sum_{n' \ne n} \alpha_{n n'}  \sigma^Z_n  \sigma^Z_{n'}  \nonumber \\
& + \dots +
\alpha_{12\dots M} \prod_{n=1}^M \sigma^Z_n \Big]
\label{bigexpansion}
\end{align}
where the coefficients are
\begin{align}
\alpha_0 & =  \frac{1}{2^M} \sum_{x \in {\cal F}}(2j_x+1 ) ,  \nonumber \\
\alpha_n & =  \frac{1}{2^M}  \sum_{x \in {\cal F}}(2j_x+1 )  (-1)^{x_n } ,  \nonumber \\
 \alpha_{n n'}  & =  \frac{1}{2^M}  \sum_{x \in {\cal F}}(2j_x +1 )  (-1)^{x_n+ x_{n'}} , \nonumber \\
\vdots& \nonumber \\
\alpha_{12\dots M} & = \frac{1}{2^M} \sum_{x \in {\cal F}}(2j_x +1 )  (-1)^{\sum_n x_n}.
\end{align}
Here the $ j_x  $ are integers that may be chosen freely. In fact it is possible to generalize (\ref{alphazcoeffs}) further if necessary, by for example adding an even integer to $ \alpha_z $. This shows that the Hamiltonian that implements the oracle has a large amount of freedom associated with it.  For our purposes the above Hamiltonian is general enough and will serve as representing a simple set of practical implementations of the oracle.

\subsection{Example and Implications}
\label{qubitexamples}

To illustrate the difference in choice of Hamiltonians, let us consider a simple example.  Let us consider the case $ M = 2 $, and a balanced oracle function such that $ f(x)= 1 $ for $\{ | 0 0 \rangle, | 11 \rangle \} $ and $ f(x)=0 $ for  $\{ | 0 1 \rangle, | 10 \rangle \} $.  Choosing $ j_x= 0 $, we have
\begin{align}
\alpha_0  & = \frac{1}{2} \nonumber \\
\alpha_n & = 0 \nonumber \\
 \alpha_{12}  & = \frac{1}{2}  .
\end{align}
The oracle Hamiltonian is thus
\begin{align}
H_f = \frac{\pi}{4}( \sigma^X_0 - 1) (1 + \sigma^Z_1 \sigma^Z_2) .
\label{hamexample1}
\end{align}
First consider operating in ``classical'' mode where the inputs of the $ x $-register is one of the logical states $ |x \rangle $ with no superposition.  For states with $ x \in {\cal F} $, we have 
$ \sigma^Z_1 \sigma^Z_2 |x \rangle = |x \rangle  $.  Then 
\begin{align}
e^{-i H_f t} | y \rangle |x \rangle  & = e^{-i \pi ( \sigma^X_0 - 1)/2} | y \rangle |x \rangle \nonumber \\
& = | \bar{y} \rangle  |x \rangle 
\end{align}
where $ \bar{y} = 1 -y $. The Hamiltonian thus flips the $y$-qubit on this cases.  For $ x \notin {\cal F} $, $ \sigma^Z_1 \sigma^Z_2 |x \rangle = -|x \rangle  $ and 
\begin{align}
e^{-i H_f t} | y \rangle |x \rangle = | y \rangle |x \rangle
\end{align}
which leaves the $y$-qubit unaffected.  Operating in ``quantum'' mode (i.e. the circuit of Fig. \ref{fig1}(b)), we have
\begin{align}
e^{-i H_f t} | - \rangle | + \rangle | + \rangle  & = e^{i \pi (1 + \sigma^Z_1 \sigma^Z_2)/2 } | - \rangle | + \rangle | + \rangle \nonumber \\
& = \sigma^Z_1 \sigma^Z_2 | - \rangle | + \rangle | + \rangle \nonumber \\
& = | - \rangle | - \rangle | - \rangle
\end{align}
where $ |\pm \rangle = (|0 \rangle \pm |1 \rangle )/\sqrt{2} $. In the above we have used the identity $ \exp( i \theta  \prod_n \sigma^Z_n ) = \cos \theta + i \sin \theta \prod_n \sigma^Z_n  $.  The Hadamard gate then rotates the above state to $ | - \rangle | 1 \rangle | 1 \rangle $, which is then measured showing that the oracle is balanced.  

A difference choice of the free parameters however can produce the same result.  Now let us choose for the same function $ j_{x=|00\rangle } = 0 $ and $ j_{x=|11\rangle } = -1 $.  This time we obtain
\begin{align}
\alpha_0  & = 0 \nonumber \\
\alpha_n & =  \frac{1}{2} \nonumber \\
 \alpha_{12}  & = 0  ,
\end{align}
which gives a Hamiltonian
\begin{align}
H_f' = \frac{\pi}{4}( \sigma^X_0 - 1) (\sigma^Z_1  + \sigma^Z_2) .
\label{hamexample2}
\end{align}
In ``classical'' mode, for $ x \in {\cal F} $, the states evolve as
\begin{align}
e^{-i H_f' t}  | y \rangle |x \rangle & = e^{ \pm i \pi( \sigma^X_0 - 1)/2} | y \rangle |x \rangle \nonumber \\
& = | \bar{y} \rangle |x \rangle 
\end{align}
which again flips the $ y $-qubit.  For cases where  $ x \notin {\cal F} $, the state remains unchanged. 
In ``quantum'' mode, we have
\begin{align}
e^{-i H_f' t}  | - \rangle  | + \rangle | + \rangle & = e^{i \pi (\sigma^Z_1  + \sigma^Z_2) /2 }   | - \rangle  | + \rangle | + \rangle \nonumber \\
& = \sigma^Z_1 \sigma^Z_2 | - \rangle  | + \rangle | + \rangle   \nonumber \\
& = | - \rangle  | - \rangle  | - \rangle  
\end{align}
We thus see that the two Hamiltonian implementations lead to the same oracle. 

It may appear curious that the Hamiltonians (\ref{hamexample1}) and (\ref{hamexample2}) lead to the same result, despite the fact that (\ref{hamexample1}) is an entangling Hamiltonian, but (\ref{hamexample2}) clearly never produces entanglement.  The reason for this are the special coefficients which for this case never result in any entanglement being generated between the qubits.  Evolving a Hamiltonian $ H = \pi \sigma^Z_1  \sigma^Z_2/2 $ applied to a state $ | + \rangle | + \rangle $ initially creates entanglement, but at the time $ t = 1 $ disentangles the qubits again.  

Based on the above result, one may speculate that perhaps it is possible to always choose an oracle Hamiltonian without any entangling terms.  This is in fact false, and is a special case for $ M \le 2 $. To see this, consider the $ M = 3 $ case and $ f(x)=1 $ for $ \{ |000 \rangle ,  |001 \rangle , |010 \rangle , |100 \rangle \} $ and $ f(x) = 0 $ otherwise. Operating in ``quantum'' mode, then from (\ref{generaloracleunitary2}) it can be seen that the oracle flips the sign of states with $ x \in {\cal F} $.  The oracle then performs the operation
\begin{align}
| + \rangle | + \rangle | + \rangle \rightarrow & \frac{1}{2} \Big( | 1 \rangle | 1 \rangle  - |0 \rangle |0 \rangle  \Big) | + \rangle \nonumber \\
& - \frac{1}{2}\Big( | 0 \rangle | 1 \rangle  + |1 \rangle |0 \rangle  \Big) | - \rangle 
\end{align}
which is obviously an entangled state.

\section{EQC Mapping Method 1: Exact approach}
\label{sec:method1}

In this section we present the first of two methods of mapping the Deutsch-Jozsa algorithm to the EQC framework.  In this method, it is possible to exactly map the qubit version of the circuit to ensembles.  This is possible for any choice of oracle, i.e. any of the freely choosable parameters in the oracle Hamiltonian.  Furthermore, the success probability is exactly 1, as in the qubit case.  The mapping however requires the preparation of Fock states, and is more susceptible to decoherence.  We later present an alternative approach that overcomes some of these issues at the expense of loss of generality of the oracle.

\subsection{Encoding}

Before introducing the quantum algorithm for Deutsch-Jozsa in EQC, we must settle on the encoding for the oracle.  In EQC, qubits are replaced by ensembles, hence there will be one ensemble which encodes the $ y $-qubit, and $ M $ ensembles encoding the $ x $-register. 

In this section we choose an encoding
\begin{align}
|0_L \rangle & \equiv | k \in \text{even} \rangle \nonumber \\
|1_L \rangle & \equiv | k \in \text{odd} \rangle 
\label{evenoddencoding}
\end{align}
where $ k \in [0,N] $ and the states on the right hand side are the Fock states as defined in (\ref{fockstatesens}) and (\ref{fockstates}). The above definition clearly has a redundancy as more than one state can encode the logical states.  This means that any one -- or superposition -- of the states that satisfy the above qualifies to be a logical state.  For example, any superposition of even $ k $ Fock states would be interpreted as a logical $ |0_L \rangle $ state. 

The above encoding is used for each of the qubits involved in the $ x $-register and the $ y $-qubit. For a given Fock state in the $ x $-register 
\begin{align}
|x \rangle =  |k_1 k_2 \dots k_M \rangle,
\label{xregisterevenodd}
\end{align}
we may obtain the logical version of this by the relation
\begin{align}
(x_n)_L = k_n \mod 2 .  
\label{decodingmodtwo}
\end{align}

Under the above encoding, we may obtain a generalized method for mapping the qubit Hamiltonians into ensemble based Hamiltonians for $ N> 1$.  Specifically we perform the mapping
\begin{align}
\sigma^Z \rightarrow S^{Z} + N + 1  .
\label{evenoddmapping}
\end{align}
To understand the origin of this mapping, consider a simple example of mapping the projection operators to the ensemble spins.  
Writing the right hand side in terms of Fock states, we have
\begin{align}
| 0 \rangle \langle 0 |  &= \frac{1}{2}(1+ \sigma^Z) \rightarrow \sum_{k=0}^N (k+1) | k \rangle \langle k |   .
\label{projectionexample}
\end{align}
Now consider that this is the Hamiltonian, and it is evolved for a particular time $ t = \pi $.  For qubits we can evaluate
\begin{align}
e^{-i \pi | 0 \rangle \langle 0 | } = | 1 \rangle \langle 1| - | 0 \rangle \langle 0 | . 
\end{align}
For the ensemble case, we have
\begin{align}
e^{-i \pi \sum_{k=0}^N (k+1) | k \rangle \langle k |} = \sum_{k \in \text{odd} } | k \rangle \langle k | - \sum_{k \in \text{even} } | k \rangle \langle k |  .
\end{align}
We thus see that under the encoding (\ref{evenoddencoding}) the effect of the mapping is the same, that it adds a negative sign for the logical 0 states, and keeps the original phase for the logical 1 states. A similar result is obtained for the logical 1 projector using
\begin{align}
| 1 \rangle \langle 1 |  & = \frac{1}{2}(1- \sigma^Z) \rightarrow \sum_{k=0}^N (-k) | k \rangle \langle k | .
\end{align}

\subsection{Oracle definition: ``classical'' mode operation}

In order that the Deutsch-Jozsa algorithm be executed in the EQC framework, we must ensure that the oracle itself can be constructed using the constraints as discussed in Sec. \ref{sec:ensembleqc}.  Specifically, we demand that the Hamiltonian is made of terms that are at most linear in total spin operators $ S^{X,Y,Z}_n $. Let us verify that the mapping (\ref{evenoddmapping}) for
oracle Hamiltonian as given by (\ref{projham}) indeed gives the desired output, working in ``classical'' mode.  Substituting, we obtain
\begin{align}
H_f & =  \frac{\pi( S^X_0 + N_0)}{2} \sum_{x' \in {\cal F}} ( 2 j_{x'} + 1) \prod_{n=1}^M \left( \frac{S^Z_n + N_n}{2} + \bar{x}_n' \right) \nonumber\\
& = \frac{\pi( S^X_0 + N_0)}{2}  \sum_{x' \in {\cal F}} ( 2 j_{x'} + 1) \nonumber \\
& \otimes \sum_{k_1 \dots k_M} (k_1 + \bar{x}_1')
\dots (k_M + \bar{x}_M') | k_1 \dots k_M \rangle \langle k_1 \dots k_M |
\label{evenoddclassical}
\end{align}
where $ \bar{x}_n' = 1 - x'_n $ and in the first line we have taken advantage of the fact that the $ j_{x'} $ can be freely chosen to absorb an appropriate factor of $ \pm 1 $ for each term in the sum. 

With the oracle operating in ``classical'' mode, the $ x $-register is prepared in a particular state (\ref{xregisterevenodd}) which represents a particular logical state according to (\ref{decodingmodtwo}). The $ y $-register can be prepared in an arbitrary state in general, lets us choose $ |0,1 \rangle \rangle $ which will illustrate the effect.  Evolving (\ref{evenoddclassical}) for $ t = 1 $ we obtain
\begin{align}
& e^{-iH_f t} |0,1 \rangle \rangle  | k_1 \dots k_M \rangle   = \nonumber \\
& \exp \left[ -i \frac{\pi}{2} ( S^X_0 + N_0)   \sum_{x' \in {\cal F}}  (2 j_{x'} + 1) \prod_{n=1}^M ( k_n + \bar{x}_n') \right] \nonumber \\
& \times |0,1 \rangle \rangle   | k_1 \dots k_M \rangle 
\end{align}
Let us first examine the parity of product for a particular term $ x' $ under the summation.  Due to the property of multiplication of odd and even integers
\begin{align}
\mbox{even} \times \mbox{even} & = \mbox{even} \nonumber \\
\mbox{even} \times \mbox{odd} &= \mbox{even} \nonumber \\
\mbox{odd} \times \mbox{odd} &= \mbox{odd} ,
\end{align}
this means that the only time that the product can evaluate to an odd integer is when
\begin{align}
x_n' = k_n \mod 2 . 
\end{align}
That is, the product is odd only when the logical state of the $ |x \rangle = | k_1 \dots k_M \rangle $ state matches the specified $ x' $.  In all other cases the product evaluates to an even integer. 

Now consider the summation over $ x' $.  There are two possible cases, either the chosen logical $ |x \rangle  $ state lies in $ \cal F $, or not.  For  $ x \notin {\cal F} $, all the terms in the sum are even, and due to the property of addition of integers
\begin{align}
\text{even} + \text{even} &= \text{even} \nonumber \\
\text{even} + \text{odd} &= \text{odd} \nonumber \\
\text{odd} + \text{odd} &= \text{even} ,
\end{align}
the sum will yield an even integer. For $ x \in {\cal F} $, then there will be exactly one term in the sum that is an odd number when $ x' = x $, and all the remaining even.  Thus the sum yields and odd number for this case.  In summary, the evolved state reduces to
\begin{align}
\exp \left[ -i \frac{\pi}{2} ( S^X_0 + N_0) P_x \right]|0,1 \rangle \rangle | k_1 \dots k_M \rangle 
\label{simplifiedclassical}
\end{align}
where 
\begin{align}
P_x = \left\{
\begin{array}{cc}
\text{odd} &  \text{if }  x \in {\cal F}  \\
\text{even} & \text{if }    x \notin {\cal F} 
\end{array}
\right. .
\label{parityfunc}
\end{align}
For $ P_x = 1$, the $ y $-ensemble is rotated by an angle $ \pi $ on the Bloch sphere, flipping its orientation. Thus for $ x \in {\cal F}  $, the $ y $-ensemble is rotated by an odd number of flips, and for $ x \notin {\cal F} $, an even number of flips:
%
%
%
\begin{align}
e^{-i H_f t} | 0,1 \rangle \rangle | x \rangle   = 
\left\{ 
\begin{array}{cc}
| 1,0 \rangle \rangle | x \rangle   &  x \in {\cal F}  \\
| 0,1 \rangle \rangle | x \rangle  &  x \notin {\cal F}  
\end{array}
\right. ,
\end{align}
up to an global phase factor. We thus see that oracle has the effect of rotating the $ y $-ensemble depending on whether the $ x $-register state is contained in $ \cal F $, which is the equivalent effect of the qubit oracle.

\subsection{``Quantum'' mode operation}

We now evaluate the operation of the oracle within the quantum circuit as shown in Fig. \ref{fig1}(b). According to the encoding (\ref{evenoddencoding}), the $ x $-register must be prepared in an even parity state, while the $ y $-ensemble must be prepared in an odd parity state. For simplicity we choose
\begin{align}
|x \rangle & = \prod_{n=1}^M | k_n = 0 \rangle = \prod_{n=1}^M | 0,1 \rangle \rangle_n \nonumber \\
|y \rangle & = | k_0 \rangle
\end{align}
where $ k_0 \in \text{odd} $.  After the Hadamard gates, this becomes
\begin{align}
|x \rangle & = \prod_{n=1}^M | \frac{1}{\sqrt{2}}, \frac{1}{\sqrt{2}} \rangle \rangle_n  \nonumber \\
|y \rangle & = | k_0 \rangle_x ,
\label{initialcondevenodd}
\end{align}
where $ | k_0 \rangle_x $ is a Fock state in the $ S^X $ basis. The $ x $-register is now in a superposition involving all states $ | k_1 \dots k_M \rangle $. 

Let us see what the effect of the Hamiltonian (\ref{evenoddclassical}) is for a general superposition state on the $ x $-register. 
Evolving for a time $ t = 1 $, we have
\begin{align}
& e^{-i H_f t} \sum_{k_1 \dots k_M} \psi_{k_1 \dots k_M} | k_0 \rangle_x  | k_1 \dots k_M \rangle = \nonumber \\
&  \sum_{k_1 \dots k_M} \psi_{k_1 \dots k_M}  \exp \left[ -i \pi k_0 \sum_{x' \in {\cal F}} (2 j_{x'} + 1) \prod_{n=1}^M ( k_n + \bar{x}_n') \right] \nonumber \\
& \times | k_0 \rangle_x  | k_1 \dots k_M \rangle  .
\end{align}
For the specific initial condition that we consider, the coefficient is
\begin{align}
\psi_{k_1 \dots k_M} =  \prod_{n=1}^M \sqrt{ \frac{1}{2^{N_n}} {N_n \choose k_n}}  .
\label{specificinit}
\end{align}
The sum in the exponent is the same quantity as that examined in the previous section, and using the property of multiplication of odd and even integers we have
\begin{align}
 e^{-i H_f t}& \sum_{k_1 \dots k_M} \psi_{k_1 \dots k_M} | k_0 \rangle_x  | k_1 \dots k_M \rangle \nonumber \\
& =  \sum_{k_1 \dots k_M} \psi_{k_1 \dots k_M} | k_0 \rangle_x  (-1)^{P_x} | k_1 \dots k_M \rangle .
\label{generalstatequantum}
\end{align}
We see that the effect of the Hamiltonian in ``quantum'' mode is to change the sign of all the terms that satisfy $ x \in {\cal F} $ under the encoding (\ref{decodingmodtwo}). 

We must now apply another Hadamard gate to the $ x $-register, which is most easily done in the spin coherent state representation. Define
the even Schrodinger cat states as
\begin{align}
| + \rangle \rangle & \equiv \frac{1}{\sqrt{2^{N}}} \sum_{k \in \text{even}} \sqrt{{N \choose k}} | k \rangle \nonumber \\
& = \frac{1}{\sqrt{2}} \left( | \frac{1}{\sqrt{2}}, \frac{1}{\sqrt{2}} \rangle \rangle + | \frac{-1}{\sqrt{2}}, \frac{1}{\sqrt{2}} \rangle \rangle \right)
\label{evencat}
\end{align}
while the odd Schr\"{o}dinger cat states are
\begin{align}
| - \rangle \rangle & \equiv \frac{1}{\sqrt{2^{N}}} \sum_{k \in \text{odd}} \sqrt{{N \choose k}} | k \rangle \nonumber \\
& = \frac{1}{\sqrt{2}} \left( | \frac{1}{\sqrt{2}}, \frac{1}{\sqrt{2}} \rangle \rangle - | \frac{-1}{\sqrt{2}}, \frac{1}{\sqrt{2}} \rangle \rangle \right) .
\label{oddcat}
\end{align}
The above was for one ensemble.  The state for a particular logical $ x $-state may then be specified according to 
\begin{align}
\prod_{n=1}^M | (-1)^{x_n} \rangle \rangle
\label{correctsector}
\end{align}
which is a superposition of states of the same parity as given in (\ref{xregisterevenodd}).  Using this notation, the state (\ref{generalstatequantum}) for the coefficients (\ref{specificinit}) can be written
\begin{align}
 e^{-i H_f t}&  | k_0 \rangle_x \prod_{n=1}^M | \frac{1}{\sqrt{2}}, \frac{1}{\sqrt{2}} \rangle \rangle_n \nonumber \\
 =  & | k_0 \rangle_x \Big( \prod_{n=1}^M | \frac{1}{\sqrt{2}}, \frac{1}{\sqrt{2}} \rangle \rangle_n   \nonumber \\
& - 2 \sum_{x \in {\cal F}} \prod_{n=1}^M | (-1)^{x_n} \rangle \rangle  \langle \langle  (-1)^{x_n} | \frac{1}{\sqrt{2}}, \frac{1}{\sqrt{2}} \rangle \rangle_n  
\Big)
\end{align}
where we have projected the parts with the specified parity in (\ref{correctsector}) and subtracted twice this in order to change the sign of these terms.  Since
\begin{align}
\langle \langle  \pm | \frac{1}{\sqrt{2}}, \frac{1}{\sqrt{2}} \rangle \rangle  =  \frac{1}{\sqrt{2}}
\end{align}
we then obtain
\begin{align}
 &e^{-i H_f t}  | k_0 \rangle_x \prod_{n=1}^M | \frac{1}{\sqrt{2}}, \frac{1}{\sqrt{2}} \rangle \rangle_n =   | k_0 \rangle_x \Big[ \prod_{n=1}^M | \frac{1}{\sqrt{2}}, \frac{1}{\sqrt{2}} \rangle \rangle_n   \nonumber \\
& - \frac{2}{2^M} \sum_{x \in {\cal F}} \prod_{n=1}^M \left( | \frac{1}{\sqrt{2}}, \frac{1}{\sqrt{2}} \rangle \rangle_n  + (-1)^{x_n} | \frac{-1}{\sqrt{2}}, \frac{1}{\sqrt{2}} \rangle \rangle_n \right) \Big] .
\label{beforehadamard}
\end{align}
Noting that there are $ 2^M/2 $ terms in the $ x $ summation, and the coefficient of $ \prod_{n=1}^M | \frac{1}{\sqrt{2}}, \frac{1}{\sqrt{2}} \rangle \rangle_n $ exactly cancels.  The set of Hadamard gates on the $ x $-register after the oracle operation finally gives the state
\begin{align}
\prod_{n=1}^M & | 0,1 \rangle \rangle_n - \frac{2}{2^M} \sum_{x \in {\cal F}} \prod_{n=1}^M \left( | 0,1 \rangle \rangle_n 
+ (-1)^{x_n} | 1,0 \rangle \rangle_n  \right) \nonumber \\
& = -\frac{2}{2^M} \sum_{z=1}^{2^M-1}  \sum_{x \in {\cal F}}  (-1)^{z\cdot x} \prod_{n=1}^M | z_n , 1- z_n \rangle \rangle  ,
\label{afterhadamard}
\end{align}
where $ z $ is an expansion index ordinarily running from $ z \in [0,2^M-1] $, in the same way as (\ref{expandinghx}). This state has exactly zero overlap with the state $ \prod_{n=1}^M  | 0,1 \rangle \rangle_n $, as the $ z=0 $ term exactly cancels, and due to $ | 0,1 \rangle \rangle $ and $ | 1,0\rangle \rangle $ being orthogonal. For a ``constant'' oracle, the overlap with $ \prod_{n=1}^M  | 0,1 \rangle \rangle_n $ is on the other hand 1.  Thus we have perfect distinguishability between the two cases and the same result for qubits has been recovered for the ensemble based method. 

The above result is the desired result in the sense that a general mapping has been obtained for an arbitrary oracle and works with (in the ideal case) probability 1.  There are some aspects which may be concerning from a practical perspective.  The first is that the odd/even encoding (\ref{evenoddencoding}) requires that one be able to prepare Fock states with a particle number resolution of 1, which can be very difficult in practice. While this may seem to make the scheme presented here unrealistic, in fact the prepared Fock states never possess any dynamics and remain static throughout both the ``classical'' and ``quantum'' circuits.  For example, in the ``classical'' circuit the $ x $-register is an eigenstate of the Hamiltonian and is unaffected by the oracle.  In ``quantum'' operation the $y$-ensemble is an eigenstate, and again remains unaffected.  Thus it would be possible to treat both these initializations classically by replacing these terms in the Hamiltonian by the desired constant. 

Another concerning aspect is that the final state involves Schrodinger cat states (\ref{evencat}) and (\ref{oddcat}).  Such states are notoriously unstable and in a realistic setting are likely to decohere very quickly. As explained in Sec. \ref{sec:ensembleqc}, in an ideal mapping from qubits to EQC, we would like to map the problem so that the decoherence is no worse than for the original qubit problem. This is however at odds with the very concept of an oracle, as it is considered to be a ``black box'' and its inner workings left unspecified. It is therefore always possible to create pathological implementations of the oracle which are highly susceptible to decoherence -- for instance one that creates a Schrodinger cat, reverses the operation to revert to the original state, then perform the oracle. Thus the emergence of Schrodinger cat states in the current encoding is the price to be paid for allowing a completely general implementation of an oracle. As we will see in the next section, some choices of the oracle implementation are better than others, when it is assumed that decoherence is present.  Thus by preferring certain oracle implementation choices, it becomes possible to implement the Deutsch-Jozsa algorithm in a more robust way.

\section{EQC Mapping Method 2: Choosable Oracle}
\label{sec:method2}

In the previous section, we presented a general mapping from the qubit version of the Deutsch-Jozsa algorithm to its EQC implementation.  While the approach has the advantage that it is completely general, it has the drawback that some undesirable decoherence-prone Schrodinger cat states are generated, and the preparation of Fock states are required. The reason that such undesirable states are involved is to accommodate a completely general oracle, 
which introduces Schrodinger cat states.  If this requirement is relaxed, then it is possible to use other encodings, which avoids some of these difficulties.

This may appear to be introducing additional assumptions into the Deutsch-Jozsa algorithm.  Nevertheless, we note that the speedup compared to the classical case is still exponential.  Consider the scenario that the particular oracle implementations for each $ f(x) $ is agreed upon initially  and chosen in a way that it satisfies --  in a suitable encoding -- the ``classical'' mode definition (\ref{oracleunitary}).  One of the oracles is then chosen without knowing which of the $ {2^M \choose 2^{M-1}} + 2 $ types of $ f(x) $ it implements.  The task is then to find whether the $ f(x) $ is constant or balanced, in the same way as the qubit Deutsch-Jozsa algorithm.  We shall see that this task can be achieved with a probability exponentially close to 1, with only one call of the oracle. Thus the main aspect of the Deutsch-Jozsa algorithm is preserved with a quantum mechanical speedup over the classical case.

\subsection{Encoding}
\label{sec:method2encoding}

The first step is again to define what the logical states that encode the inputs and outputs of the oracle are.  In this approach, we define logical states with an analogous state on the Bloch sphere for the ensembles as for qubits.  Hence the logical states for $ y \in \{0,1 \} $ have a correspondence
\begin{align}
| y \rangle \leftrightarrow | y \rangle \rangle \equiv | y, 1-y \rangle \rangle
\label{ymapping}
\end{align}
for the $ y $-qubit, which is now a $ y$-ensemble, and for the $ x $-register we have
\begin{align}
|x \rangle & \equiv | x_1 \rangle | x_2 \rangle \dots | x_M \rangle \nonumber \\
&  \updownarrow \nonumber \\
| x \rangle \rangle & \equiv | x_1, 1-x_1 \rangle \rangle | x_2, 1-x_2 \rangle \rangle \dots | x_M, 1-x_M \rangle \rangle ,  
\label{xregistermapping}
\end{align}
where we assume $x_n \in \{ 0,1 \} $. 

For this encoding, the Pauli operators are mapped according to 
\begin{align}
\sigma^Z \rightarrow \frac{S^Z}{N} .
\label{secondmapping}
\end{align}
The normalization with $ N $ means that when the mapped operator acts on a state
\begin{align}
\frac{S^Z}{N} | k \rangle = \left( \frac{2k}{N} - 1 \right) | k \rangle
\end{align}
where $ k \in [0,N] $, so that for the extremal states $ | k = 0 \rangle = | 0, 1 \rangle \rangle $ and $ | k = N \rangle = | 1,0 \rangle \rangle $, 
\begin{align}
\frac{S^Z_n}{N_n} | x \rangle \rangle = (\pm 1)^{x_n} | x \rangle \rangle ,
\end{align}
which is identical to the qubit case.

\subsection{Oracle definition: ``classical'' mode operation}

We first write the effect of the oracle working in ``classical'' mode, under the encodings (\ref{ymapping}) and (\ref{xregistermapping}).  
From the qubit definition (\ref{oracleunitary}), writing the ensembles explicitly a valid oracle for EQC must satisfy
\begin{align}
U_f  | y \rangle \rangle | x \rangle \rangle  &  = U_f  | y,1-y \rangle \rangle \prod_{n=1}^M   | x_n, 1-x_n \rangle \rangle \nonumber \\
 &  =  | y \oplus f(x) ,1- y \oplus f(x) \rangle \rangle  | x \rangle \rangle \nonumber \\
& = | y \oplus f(x) \rangle \rangle | x \rangle \rangle
\label{becoracledef}
\end{align}
We note that the above definition only constrains the states $ | x_n, 1-x_n \rangle \rangle $ and $ | y, 1- y \rangle \rangle $, where $ x_n, y \in \{0,1 \} $.  These are only 2 states out of $ N + 1 $ states per ensemble, hence this clearly leaves a lot of states unspecified.  This is in practice not a problem as we will see below, as only linear powers of the total spin operators $ S^{X,Y,Z}_n $ are used in the mapping which has the effect of defining the remaining states by linearly interpolating between the definitions. 

First considering the constant cases, for $ f = 0 $ we have from (\ref{fzeroham})
\begin{align}
H_{f=0} = 0 ,
\end{align}
where we have chosen the free parameter $ j = 0 $, as in this approach the assumption is that we are free to choose the most convenient implementation of an oracle.  For $ f = 1 $, since rotations of a single ensemble have identical time coefficients as qubits, this suggests that we have 
\begin{align}
H_{f=1} = \pi \left( \frac{S^X_0 - N_0}{2} \right).  
\label{foneeqc}
\end{align}
where we have chosen $ 2j+1 = N_0 $. While in the qubit case $ 2j+1 $ is required to be an odd integer, in this case it is unnecessary and (\ref{foneeqc}) reproduces the desired oracle (\ref{becoracledef}) for any $ N_0 $.  

For the balanced cases, the qubit Hamiltonian has the form given in (\ref{projham}). The sum in this expression evaluates to an odd integer if the Hamiltonian operates on a state with $ x \in {\cal F} $, and an even integer for $ x \notin {\cal F} $. The oracle thus flips the $ y $-input conditionally on the $ x $-register. This same logic is preserved under the mapping (\ref{secondmapping}), which leads us to the Hamiltonian
\begin{align}
H_f & = \pi \left(\frac{S^X_{0} - N_0}{2} \right) \nonumber \\
& \otimes \sum_{x \in {\cal F}} (2 j_x +1 ) \prod_{n=1}^M \frac{1}{2} \left[ 1 + (-1)^{x_n} \left(\frac{S^Z_n}{N_n} \right) \right] .  
\end{align}
Following the same steps as the qubit case to derive the expanded version of the EQC oracle Hamiltonian, we obtain
\begin{align}
& H_f = \pi \left(\frac{S^X_{0} - N_0}{2} \right)  \sum_{z=0}^{2^M-1}\alpha_z \prod_{n=1}^M \left(\frac{S^Z_n}{N_n} \right)^{z_n}  . 
\label{expandedbecham}
\end{align}
Evolving this Hamiltonian for a time $ t = 1 $, this satisfies the oracle definition (\ref{becoracledef}).  We note that this definition satisfies the constraints of EQC, that the Hamiltonian can be written entirely in terms of linear products of the total spin operator on each ensemble.  While we have not yet chosen the free parameters $ j_x $ which fix $ \alpha_z $, we shall see in the next section that there is a convenient choice which simultaneously simplifies the implementation and avoids decoherence-prone states. 


\subsection{``Quantum'' mode operation}

Let us now observe what the effect of the oracle is when applied in the Deutsch-Jozsa circuit as shown in Fig. \ref{fig1}(b).  The $ x $-register first starts in the state $ | 0 \rangle \rangle $, while the $ y $-ensemble starts in the state $ | 1 \rangle \rangle $.  After the Hadamard gates are applied, the state becomes
\begin{align}
| \psi_{\text{init}} \rangle = |  \frac{1}{\sqrt{2}} , \frac{-1}{\sqrt{2}} \rangle \rangle_0 \prod_{n=1}^M | \frac{1}{\sqrt{2}} ,\frac{1}{\sqrt{2}} \rangle \rangle_n .  
\end{align}
The $ y $-ensemble state is an eigenstate of the $ S^X_0 $ operator, hence for constant oracles this leave the registers unchanged up to a phase: 
\begin{align}
e^{-i H_{f=0} t } | \psi_{\text{init}} \rangle = | \psi_{\text{init}} \rangle \nonumber \\
e^{-i H_{f=1} t } | \psi_{\text{init}} \rangle = (-1)^{N_0} | \psi_{\text{init}} \rangle ,
\end{align}
where we evolve for a time $ t = 1 $. 

For the balanced case Hamiltonian, we have (\ref{expandedbecham}), or in expanded form we may write
\begin{align}
& e^{-i H_f t} | \psi_{\text{init}} \rangle  =  \exp \Big[ i \pi \Big( N_0 \alpha_0
+ \sum_n \frac{\alpha_n N_0 }{N_n} S^Z_n  \nonumber \\
& + \sum_n \sum_{n' \ne n} \frac{\alpha_{n n'} N_0}{N_n N_{n'}}  S^Z_n  S^Z_{n'}  +  \dots +
\frac{\alpha_{12\dots M} N_0} {\prod_{n=1}^M N_n}  \prod_{n=1}^M S^Z_n \Big) \Big] | \psi_{\text{init}} \rangle 
\label{becbigexpansion}
\end{align}
where the same steps leading to (\ref{bigexpansion}) were performed in this case. It is clear that in the Deutsch-Jozsa circuit, for the constant cases the Hamiltonians leaves the $ x $-register unaffected.  Meanwhile, in the balanced cases the Hamiltonian involves a polynomial in $ S^Z_n $ operators.  In order to distinguish between the constant and balanced cases, what is required is that the $ S^Z_n $ terms rotate  $ | \psi_{\text{init}} \rangle $ sufficiently far away such that it is an orthogonal state.  Once it is rotated to an orthogonal state, it should be discriminable via the measurement state at the end of the gate sequence.

\begin{figure}
\includegraphics[width=\columnwidth]{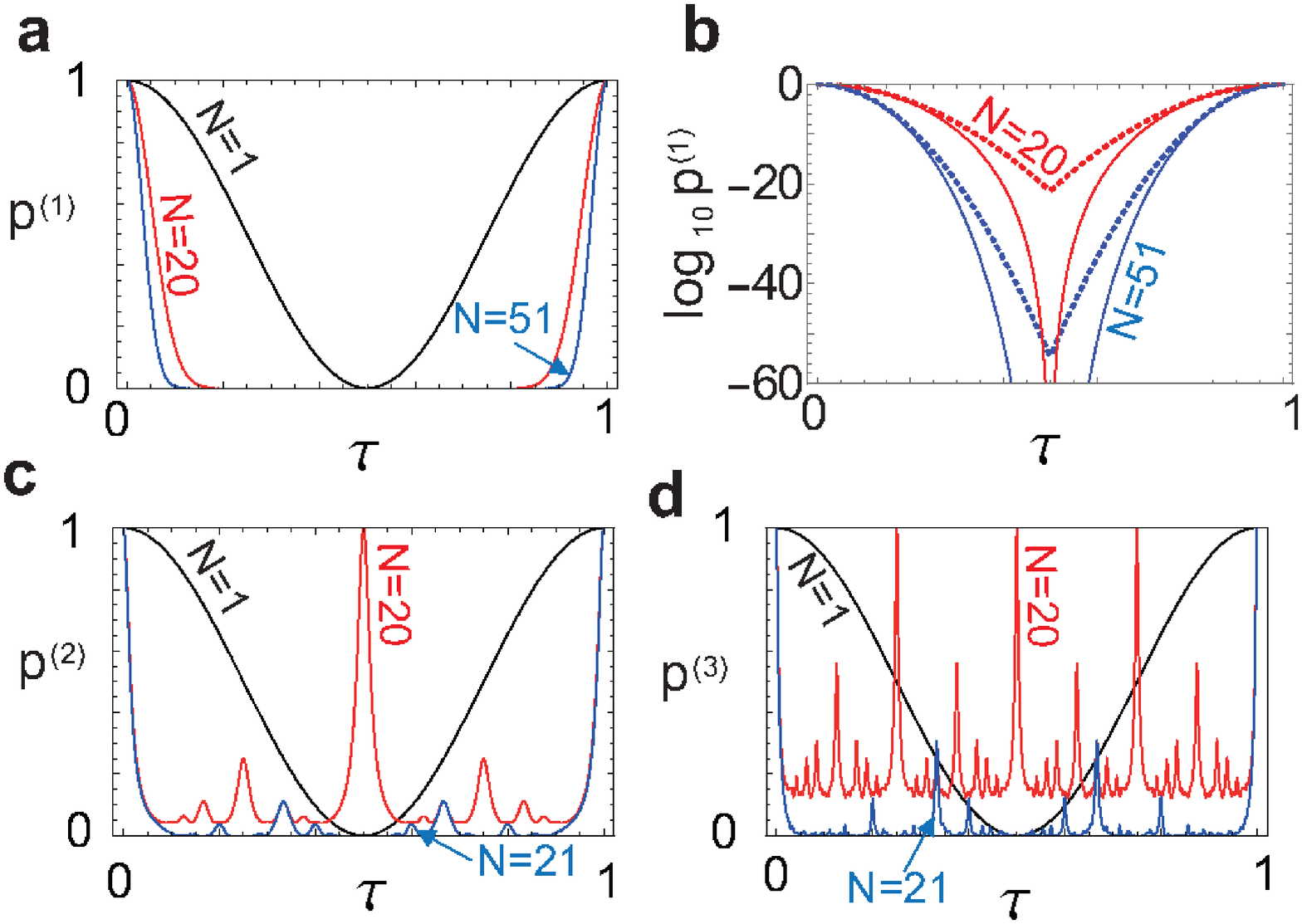}
\caption{The probability $ p^{(m)} $ of remaining in the initial state $ | \psi_{\text{init}} \rangle $ after evolving with a Hamiltonian $ \pi \prod_{n=1}^m S^Z_n $, as defined in (\ref{probfunction}).  (a) $ m = 1 $ with $ N_1 = N $ as marked. (b) $ m = 1 $ on a logarithmic scale with $ N_1 = N $ as marked (solid lines), with the approximation (\ref{approxpone}) (dashed lines). (c) $ m = 2 $ with $ N_1=N_2= N $ as marked. (d) $ m = 3 $ with $ N_1=N_2=N_3= N $ as marked.  }
		\label{fig2}
\end{figure}

To see to what extent the various terms in the expansion (\ref{becbigexpansion}) take the $ x $-register away from its initial state, let us compute the overlap probability 
\begin{align}
p^{(m)} (\tau) & =  \left|  \langle \psi_{\text{init}} | e^{i \pi \tau \prod_{n=1}^m S^Z_n } | \psi_{\text{init}} \rangle \right|^2  .
\label{probfunction}
\end{align}
This represents the probability that the initial state $ | \psi_{\text{init}}  \rangle $ remains in the same state after evolving with various terms in the expansion (\ref{becbigexpansion}).  For  balanced Hamiltonians, ideally this is zero such that the final detection probability of $ | x = 0 \rangle \rangle $ is zero.  Here $ \tau $ is a parameter which represents the coefficient of $ \prod_{n=1}^m S^Z_n $ up to a factor of $\pi$.  The first few expressions may be evaluated by expanding the coherent states into Fock states, we write the results below:
\begin{align}
\label{p1formula}
p^{(1)} (\tau) & = \cos^{2 N_1} ( \pi \tau)  \\
p^{(2)} (\tau) & = \frac{1}{4^{N_1}} \left|  \sum_{k_1} {N_1 \choose k_1} \cos^{N_2} \left[ \pi \tau (2 k_1 -N_1)  \right]  \right|^2  \\
p^{(3)} (\tau) & =  \frac{1}{4^{N_1+N_2}}  \Big|  \sum_{k_1 k_2} {N_1 \choose k_1} {N_2 \choose k_2}  \nonumber \\
& \times \cos^{N_3} \left[ \pi \tau (2 k_1 -N_1) (2 k_2 -N_2)  \right]  \Big|^2 .
\end{align}
The probabilities are plotted in Fig. \ref{fig2}. We see that all the plots are periodic with period $ \tau = 1 $.  For the qubit case $ N = 1 $, all curves give the same behavior, where the probability is zero at $ \tau = 1/2 $.  For $ N> 1 $, in general we see more complex  behavior where $ p^{(m)} (\tau = 1/2 ) $ is not necessarily equal to zero. There is a strong even/odd dependence to the curves where qualitatively different behavior is seen for each case.  In particular, for $ N_n $ all odd the curves have a zero at $ \tau = 1/2 $, however when even $ N_n $ are involved this can instead become 1.  For $ m \ge 2 $ the $ N_n $ even cases do not possess a zero at all for any time. The exception to this complex behavior is the $ m = 1 $ case, where there is no even/odd effect, and for any $ N_1 $ we have $ p^{(1)} (\tau = 1/2) = 0 $, as is easily seen from (\ref{p1formula}).  In fact for this case we may approximate for large $ N_1 $
\begin{align}
p^{(1)} (\tau) \approx \sum_j e^{-N_1 \pi^2 (\tau + j)^2},
\label{approxpone}
\end{align}
where $ j $ are integers. This approximation is in valid in the region where the probability is non-negligible, and  $N_1 \gg 1$  as shown  Fig. \ref{fig2}(b).  Due to the factor of $ N_1 $ in the Gaussian, for larger $ N_1 $ it is in fact very easy to suppress the overlap probability to zero for the $ m = 1 $ case.  For the qubit case, a time of exactly $ \tau = 1/2 $ to suppress the probability, whereas for large ensembles we have a window of $ 1/\sqrt{N_1} \lesssim \tau \lesssim  1- 1/\sqrt{N_1} $.  

This suggests that in terms of minimizing the overlap probability it is most effective to use the $ m = 1 $ term, as it gives a strong suppression and is most predictable with respect to the number of atoms.  The higher order terms would require control of the number of particles in the ensemble to within one atom to control the parity, and is far less desirable.  Fortunately, as discussed in Sec. \ref{sec:qubitoracle}, it is always possible to choose the oracle in a way such that it contains at least one of the $ m= 1 $ terms with the desired coefficient of $ \tau = 1/2 $.  It is an arbitrary choice of which ensemble to have the $ \tau = 1/2 $ coefficient, here we shall choose $ n = 1 $.  Returning to the coefficients defined in (\ref{bigexpansion}), consider making the choice
\begin{align}
j_x = - x_1.
\label{magicchoice}
\end{align}
For this choice the coefficients are
\begin{align}
\alpha_0 & = \frac{1}{2^M} \sum_{x \in {\cal F}}(-1)^{x_1} ,  \nonumber \\
\alpha_n & =  \frac{1}{2^M}  \sum_{x \in {\cal F}} (-1)^{x_n+ x_1 } ,  \nonumber \\
 \alpha_{n n'}  & =   \frac{1}{2^M}  \sum_{x \in {\cal F}}(-1)^{x_n+ x_{n'}+ x_1} , \nonumber \\
\vdots& \nonumber \\
\alpha_{12\dots M} & =  \frac{1}{2^M} \sum_{x \in {\cal F}} (-1)^{\sum_{n=2}^M x_n}.
\label{choicecoeffs}
\end{align}
Specifically,  this choice gives
\begin{align}
\alpha_1 = \frac{1}{2} . 
\end{align}
Thus by this particular choice of $j_x $, we have ensured that  the coefficient of the $ S^Z_1 $ term in (\ref{becbigexpansion}) is equal to 
\begin{align}
-\frac{\pi N_0 }{2 N_1} .
\end{align}
To ensure that the $n = 1 $ ensemble in the $ x $-register is orthogonal, this coefficient suggests that we should have $ N_0 = N_1 $, such that
\begin{align}
e^{i \pi S^Z_1/2} & \prod_{n=1}^M | \frac{1}{\sqrt{2}},  \frac{1}{\sqrt{2}} \rangle \rangle_n \nonumber \\
& = e^{-i \pi N_1/2}  | \frac{-1}{\sqrt{2}},  \frac{1}{\sqrt{2}} \rangle \rangle_1  \prod_{n=2}^M | \frac{1}{\sqrt{2}},  \frac{1}{\sqrt{2}} \rangle \rangle_n ,
\label{firststep}
\end{align}
which has zero overlap with the initial state. However, this is not a very sensitive requirement as
for large $ N_1 $ the overlap quickly vanishes as seen in Fig. \ref{fig2}(b).  Thus to a reasonable approximation having 
\begin{align}
2 \sqrt{N_1} \lesssim N_0 \lesssim  2 N_1 .
\end{align}
should give a sufficiently low overlap state. 

The choice (\ref{magicchoice}) fixes the coefficient of $ S^Z_1 $, but also affects all the other coefficients (\ref{choicecoeffs}).  How can we be sure that the other coefficients do not spoil the orthogonality that is created by the $ S^Z_1 $ term? To see this first note that all the terms in (\ref{becbigexpansion}) commute, so that we may apply any of the terms in any order. Applying the $ S^Z_1 $ term first, then what we require is that the remaining terms  (\ref{becbigexpansion}) do not somehow make (\ref{firststep}) again have an overlap with the initial state $ | \psi_{\text{init}} \rangle $.  For the other first order $ m = 1 $ terms, this does not affect the $ n = 1 $ ensemble, as it they rotate the other coherent states in (\ref{firststep}) away from $ | \frac{1}{\sqrt{2}}, \frac{1}{\sqrt{2}} \rangle \rangle $.  For the higher order terms $ m \ge 2 $, we observe that all the coefficients are bounded by 
\begin{align}
| \alpha_z | = \left| \frac{1}{2^M} \sum_{x \in {\cal F}}  (-1)^{z \cdot x + x_1} \right| \le \frac{1}{2} .
\end{align}
Assuming that all the ensembles are approximately of the same size $ N_n \approx N $, then according to (\ref{becbigexpansion}) the coefficient of an $ m $th order term is 
\begin{align}
\tau \sim \frac{\alpha_z}{N^{m-1}}.
\label{typicaltimes}
\end{align}
Thus for this choice of $ j_x $, the coefficients diminish for higher orders. This suggests that the higher order $ m \ge 2 $ terms may be negligible, in particular for large $ N $.   

\begin{figure}
\includegraphics[width=\columnwidth]{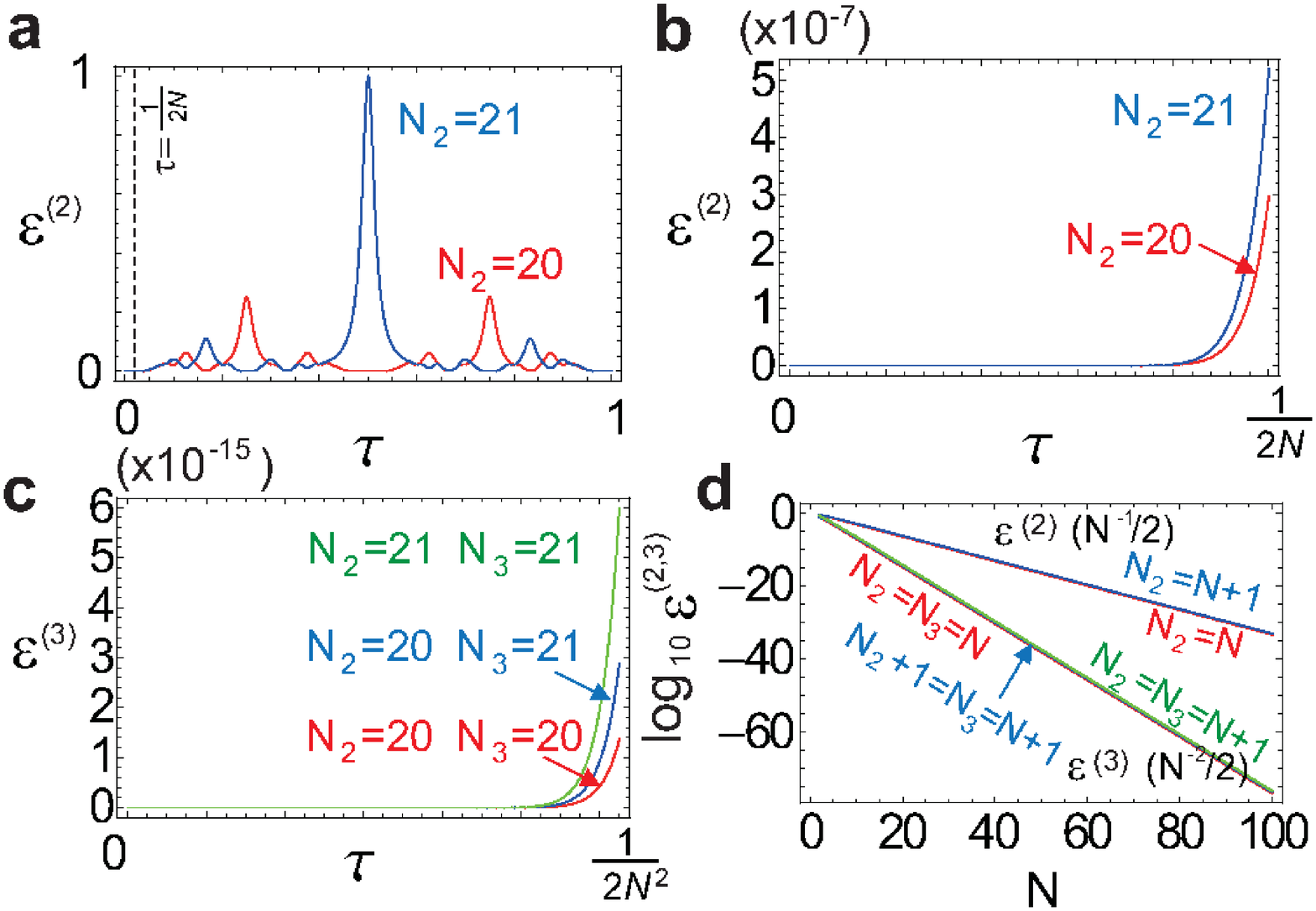}
\caption{Error probabilities as defined in (\ref{errordefinition}).  This corresponds to the probability for the measurement yielding $ | \psi_{\text{init}} \rangle $ for a balanced function for various order terms, after the first order term is applied. (a)(b) The second order error probability $ m = 2 $ for $ N_1 = 20 $ and $ N_2 $ as marked. (c) The third order error probability $ m = 3 $, for $ N_1=20 $ and $ N_2, N_3 $ as marked. (d) The second $ m = 2 $ at time $ \tau = 1/2N $ and third $ m = 3 $ error probabilities at time $ \tau = 1/2N^2 $ as a function of the particle number $ N $.  The particle numbers are set as marked, and $ N_1 = N $.  }
		\label{fig3}
\end{figure}

To verify this, let us calculate explicitly the effect of whether the higher order terms spoil the orthogonality initially created by $ S^Z_1 $. Consider the
following probability function which measures how well the orthogonality is preserved after $ S^Z_1 $ initially creates an orthogonal state:
\begin{align}
\varepsilon^{(m)} (\tau) & =  \left|  \langle \psi_{\text{init}} | e^{i \pi \tau \prod_{n=1}^m S^Z_n } e^{i \pi S^Z_1 /2}  | \psi_{\text{init}} \rangle \right|^2  .
\label{errordefinition}
\end{align}
As with (\ref{probfunction}), we would like this to be as close to zero as possible.  This can be evaluated to be
\begin{widetext}
\begin{align}
\varepsilon^{(2)} (\tau) & = \left\{
\begin{array}{cl}
\frac{1}{4^{N_2}} \left|  \sum_{k_2=0}^{N_2} {N_2 \choose k_2} \sin^{N_1} \left[ \pi \tau (2 k_2 -N_2)  \right]  \right|^2  & \text{ if } N_1 \in \text{even} \\
0 & \text{ if }  N_1 \in \text{odd}
\end{array}
\right.
\nonumber \\
\varepsilon^{(3)} (\tau) & = \left\{
\begin{array}{cl}
\frac{1}{4^{N_2+N_3}}  \Big|  \sum_{k_2=0}^{N_2} \sum_{k_3=0}^{N_3} {N_2 \choose k_2} {N_3 \choose k_3}  \sin^{N_1} 
\left[ \pi \tau (2 k_2 -N_2) (2 k_3 -N_3)  \right] \Big|^2   & \text{ if } N_1 \in \text{even} \\
0 & \text{ if }  N_1 \in \text{odd}
\end{array}
\right.
\label{explicitpminus}
\end{align} 
\end{widetext}
For the case that $ N_1 $ has an odd number of particles the probability is exactly zero as the summands in (\ref{explicitpminus}) are odd functions. Similarly to $ p^{(m)} $, the above functions have a strong dependence on whether the other ensembles involved have an even or odd number of particles.  Fig. \ref{fig3}(a) shows the large timescale behavior for $ m=2 $. We see that the functions do possess multiple zeros for both even and odd $ N_2 $, which is in contrast to $ p^{(2)} $, where no zeros are present for even $ N_1, N_2 $.  The relevant timescale for our choice of $ j_x $ is (\ref{typicaltimes}), which we plot in Figs. \ref{fig3}(b)(c).  We see that both for the $ m = 2 $ and $ m =3 $ cases the probability remains extremely small, at the $ \sim 10^{-7} $ and $ \sim 10^{-15} $ levels respectively even for the maximal case where $ | \alpha_z | = 1/2 $.  In Fig. \ref{fig3}(d) we show the particle number dependence of the probability at the maximal case of (\ref{typicaltimes}) on a semi-logarithmic plot. The odd/even dependence gives only a minor variation on this scale, and follow a simple exponential form.  A fit of the data gives the following estimate of the probability
\begin{align}
\varepsilon^{(2)} (- \frac{1}{2N} < \tau < \frac{1}{2N} ) & \lesssim e^{0.81-0.77 N} \nonumber \\
\varepsilon^{(3)} ( - \frac{1}{2N^2} < \tau < \frac{1}{2N^2} ) & \lesssim e^{2.62 - 1.78 N} .
\label{fitscaling}
\end{align}

Another source of potential errors is due to the variations in the particle number between the ensembles, which we have so far assumed that $ N_n = N $.  The effect of different particle numbers in the ensembles is to modify the coefficients in (\ref{becbigexpansion}).  Assuming that the ensembles can be prepared within $ \sim 10 \% $, this has the effect of shifting $\tau $ by this factor, which will generally have the same behavior as (\ref{fitscaling}).  A potentially more serious effect is an imperfect rotation of the $ S^Z_1 $ ensemble, which is the primary source of the desired orthogonality.  Such imperfect rotations can be described by
\begin{align}
\varepsilon^{(1)} ( \tau) & =  \sin^{2 N_1} ( \pi \tau) \nonumber \\
& \approx (\pi \tau)^{2 N_1}= e^{ 2 \ln ( \pi \tau) N_1 } .
\label{errorone}
\end{align}
For ensembles prepared within $ \sim 10 \% $, this corresponds to an additional rotation of $ \tau = 0.05 $, which gives an exponent $ \varepsilon^{(1)} ( \tau) \sim e^{-3.7 N } $.  This is in fact suppressed more than the error contributions of (\ref{fitscaling}).  

We thus conclude that the largest error contribution is $ \varepsilon^{(2)} $, due to the second order terms $ m = 2 $. This is reasonable as these have the largest coefficients after the linear terms, which are the desired terms.  In all cases the probability of obtaining the original state is exponentially suppressed with the particle number. In realistic systems the number of particles with $ N \gtrsim 10^3 $, the above estimates would give an error probability that is completely negligible ($ \varepsilon^{(2)} \sim 10^{-334} $). Thus at least in the ideal case,  the above shows that it is possible distinguish constant and balanced oracles in the same way as the standard Deutsch-Jozsa algorithm, i.e. by detection, or lack of detection, respectively of $ | x=0 \rangle \rangle $. While the probability of obtaining $ | x=0 \rangle \rangle $ is not strictly zero in the balanced case, it is highly suppressed for reasonable parameters, to the extent that it is negligible.

\section{Examples}

In this section we present some explicit examples of EQC implementations of the Deutsch-Jozsa algorithm.

\subsection{Deutsch's algorithm}

In the case of $ M = 1 $, the Deutsch-Jozsa algorithm reduces to Deutsch's algorithm. In this case it is in fact possible to use the encoding presented in Sec. \ref{sec:method2encoding} to obtain a mapping which works in the EQC framework with negligible error, even for a generalized oracle.   

In the case of Deutsch's algorithm, there are only four possible $ f(x) $, with two constant and two balanced.  The Hamiltonians corresponding to each case are as follows
\begin{align}
H_{f=0} & = 2 \pi j \\
H_{f=1} & = \pi (2j'+1) \left( \frac{S^X_0 - N_0}{2} \right) \\
H_{f=\{1,0\}} & = \pi (2 j_0+1) \left( \frac{S^X_0 - N_0}{2} \right) \left(\frac{1+ S^Z_1/N_1}{2} \right) \\
H_{f=\{0,1 \}} & = \pi (2 j_1+1) \left( \frac{S^X_0 - N_0}{2} \right) \left(\frac{1- S^Z_1/N_1}{2} \right)
\end{align}
where $ j, j', j_1, j_0 $ are integers that may be chosen freely. 

For ``classical'' operation, the above Hamiltonians satisfy the requirements of a valid oracle under the encoding (\ref{ymapping}) and (\ref{xregistermapping}). Evolving the above Hamiltonians for $ t =1 $, and using $ S^Z_1/N_1 | x \rangle \rangle = (-1)^x |  x \rangle \rangle  $, we have
\begin{align}
e^{-i H_{f=0} t} |x \rangle \rangle | y \rangle \rangle & = |x \rangle \rangle | y \rangle \rangle \nonumber \\
e^{-i H_{f=1} t} |x \rangle \rangle | y \rangle \rangle & =  |x \rangle \rangle | \bar{y} \rangle \rangle 
\nonumber \\
e^{-i H_{f=\{1,0\}} t} |0 \rangle \rangle | y \rangle \rangle & =  |0 \rangle \rangle | \bar{y} \rangle \rangle 
\nonumber \\
e^{-i H_{f=\{1,0\}} t} |1 \rangle \rangle | y \rangle \rangle & =  |1 \rangle \rangle | y \rangle \rangle 
\nonumber \\
e^{-i H_{f=\{0,1\}} t} |0 \rangle \rangle | y \rangle \rangle & =  |0 \rangle \rangle | y \rangle \rangle 
\nonumber \\
e^{-i H_{f=\{0,1\}} t} |1 \rangle \rangle | y \rangle \rangle & =  |1 \rangle \rangle | \bar{y} \rangle \rangle  ,
\end{align}
where $ \bar{y} = 1- y $ and we have discarded any irrelevant global phase factors. 

In ``quantum'' mode, after the initial Hadamard gates, the Hamiltonian is applied on the state
\begin{align}
 | \psi_{\text{init}} \rangle = | \frac{-1}{\sqrt{2}}, \frac{1}{\sqrt{2}} \rangle \rangle_0 | \frac{1}{\sqrt{2}}, \frac{1}{\sqrt{2}} \rangle \rangle_1 .
\end{align}
For the constant cases, the Hamiltonians clearly leave the ensemble $ n =1 $ untouched, so the probability of obtaining $ | 0 \rangle \rangle $ at the measurement is 1. For the balanced cases, the states evolve as
\begin{align}
& e^{-i H_{f=\{1,0\}} t }  | \psi_{\text{init}} \rangle = e^{i \frac{N_0 \pi (2 j_0+1)}{2N_1} S^Z_1 t} 
| \psi_{\text{init}} \rangle \nonumber \\
& = | \frac{-1}{\sqrt{2}}, \frac{1}{\sqrt{2}} \rangle \rangle_0
| \frac{e^{i \frac{\pi N_0  (2 j_0+1)}{2N_1}} }{\sqrt{2}}, \frac{e^{-i \frac{\pi N_0  (2 j_0+1)}{2N_1}} }{\sqrt{2}} \rangle \rangle_1  \nonumber \\
\end{align}
and 
\begin{align}
& e^{-i H_{f=\{0,1\}} t }  | \psi_{\text{init}} \rangle = e^{-i\frac{N_0 \pi (2 j_1+1)}{2N_1}S^Z_1 t} | \psi_{\text{init}} \rangle \nonumber \\
& = | \frac{-1}{\sqrt{2}}, \frac{1}{\sqrt{2}} \rangle \rangle_0
| \frac{e^{-i \frac{\pi N_0  (2 j_1+1)}{2N_1}} }{\sqrt{2}}, \frac{e^{i \frac{\pi N_0  (2 j_1+1)}{2N_1}} }{\sqrt{2}} \rangle \rangle_1  .
\end{align}
The probability of obtaining the initial state is thus
\begin{align}
|\langle \psi_{\text{init}} | e^{-i H_{f=\{1,0\}} t } | \psi_{\text{init}} \rangle|^2 & = 
\cos^{2 N_1}  ( \pi N_0  (2 j_0+1)/2N_1 ) \nonumber \\
|\langle \psi_{\text{init}} | e^{-i H_{f=\{0,1\}} t } | \psi_{\text{init}} \rangle|^2 & = 
\cos^{2 N_1}  ( \pi N_0  (2 j_1+1)/2N_1 )  .
\label{probdeutsch}
\end{align}
For $ N_0=N_1 $, the right hand side evaluates to exactly zero for all $ j_0, j_1 $.  For $ N_0 \ne N_1 $, the probability depends upon the particular choice of free parameters $ j_0, j_1 $.  In order to avoid amplifying the particle number mismatch between the ensembles, the safest choice is $ j_0=j_1 = 0 $.  Assuming that $ N_0 \approx N_1 $, then in a similar way to (\ref{errorone}) we can estimate the  probability to be
\begin{align}
p \sim \left( \frac{\pi ( 1 - N_0/N_1)}{2} \right)^{2 N_1}
\end{align}
which is a very small number for typical parameters.  For example, for $N_0=1000, N_1=1100  $, one obtains $ p \sim 10^{-1863} $, which is negligible.  Thus a constant or balanced oracle can be distinguished by a measurement of $ | x= 0 \rangle \rangle $, in exactly the same way as the qubit version of Deutsch's algorithm.

\begin{table*}[t]
\centering
\begin{tabular}{|c|cc|cccccc|}
\hline
$x$  & $ x_2 $ & $ x_1 $ & $f_1(x)$ &$f_2(x)$ & $f_3(x)$ & $f_4(x)$ & $f_5(x)$ &$f_6(x)$ \\
\hline
0& 0& 0 & 0          & 0          & 0          & 1          & 1          & 1          \\
1& 0& 1& 0         & 1          & 1          & 0          & 0          & 1          \\
2& 1& 0 & 1          & 0          & 1          & 0          & 1          & 0          \\
3& 1& 1 & 1          & 1          & 0          & 1          & 0          & 0          \\ 
\hline
\end{tabular}
\caption{Balanced Deutsch-Jozsa functions for $ M = 2 $. }
\label{tbl:n=2}
\end{table*}

\subsection{M=2 case, Method 2}

For $ M =2 $ there are 6 types of balanced oracles as shown in Table \ref{tbl:n=2}.  Of the balanced oracles, cases 3 and 4 are most non-trivial as they have a dependence on both of the input parameters $ x_1 $ and $ x_2 $.  For other cases, the functions are independent of one of the variables (for example, $ f_1(x) $ is independent of $ x_1 $) and give simpler results. Cases 3 and 4 are only different by a global negation, hence we will focus on case 4 -- which is the same as that examined in Sec. \ref{qubitexamples} -- for this section.

As seen from (\ref{hamexample1}) and (\ref{hamexample2}), there is not a unique way to realize the oracle corresponding to this (or any) function.  In the case that we are allowed to choose the oracle implementation, the simpler choice would be (\ref{hamexample2}), which only involves linear terms in the Pauli operators for the $ x $-register. In this case we would follow the procedure in Sec. \ref{sec:method2}.  The Hamiltonian in this case would be 
\begin{align}
H_{f_4} = \frac{\pi}{2} \left( \frac{S^X_0 - N_0}{2} \right) \left( \frac{S^Z_1}{N_1} + \frac{S^Z_2}{N_2} \right) .
\end{align}
Encoding the logical states as in Sec. \ref{sec:method2encoding}, and operating in ``classical'' mode, 
\begin{align}
e^{-i H_{f_4} t} &  | y \rangle \rangle | x \rangle \rangle = \left\{
\begin{array}{cc}
 | y \rangle \rangle | x \rangle \rangle & x= 1,2 \\
 | \bar{y} \rangle \rangle | x \rangle \rangle & x= 0,3 \\
\end{array}
\right. ,
\end{align}
up to an irrelevant global phase.  In ``quantum'' mode, the initial state evolves to
\begin{align}
e^{-i H_{f_4} t} | \psi_{\text{init}} \rangle = | \frac{-1}{\sqrt{2}}, \frac{1}{\sqrt{2}} \rangle \rangle_0
| \frac{e^{-i\pi \frac{ N_0}{N_1}} }{\sqrt{2}}, \frac{1}{\sqrt{2}} \rangle \rangle_1 
| \frac{e^{-i\pi \frac{ N_0}{N_2}} }{\sqrt{2}}, \frac{1}{\sqrt{2}} \rangle \rangle_2 ,
\end{align}
which is orthogonal to $ | \psi_{\text{init}} \rangle $ for $ N_0 = N_1 = N_2 $. Similar probability expressions to (\ref{probdeutsch}) case can be evaluated.

\subsection{M=2 case, Method 1}
\label{sec:decoherence}

Let us also take the approach of Sec. \ref{sec:method1} to implement the $ M=2 $ Deutsch-Jozsa algorithm in EQC. One of the drawbacks of this method is that it generates Schrodinger cat states which are vulnerable to decoherence.  We calculate the performance under dephasing to analyze the sensitivity of the scheme to decoherence. 
 
Following the exact mapping procedure as discussed in Sec. \ref{sec:method1}, and substituting (\ref{evenoddmapping}) into (\ref{hamexample1}), we obtain
\begin{align}
H_{f_4} = \frac{\pi}{4} \left( S^X_0 + N_0 \right)  \left( 1 + (S^Z_1+N_1 +1)(S^Z_2+N_2 +1) \right)
\end{align}
Operating in ``classical'' mode, consider evolving the above Hamiltonian on the state (\ref{xregisterevenodd}), which gives
\begin{align}
& e^{-i H_{f_4} t}  | 0,1 \rangle \rangle | k_1 k_2 \rangle  = \nonumber \\
& \exp \left[ -i \frac{\pi}{2} (S^X_0 + N_0 ) \left( 2 k_1 k_2 + k_1 + k_2 + 1 \right) \right]
 | 0,1 \rangle \rangle | k_1 k_2 \rangle  .
\label{classicalexample}
\end{align}
The factor $ 2 k_1 k_2 + k_1 + k_2 + 1  $ can be observed to be an odd integer when both $ k_1 $ and $ k_2 $ are odd or even, and is an even integer when one of $ k_1 $ and $ k_2 $ are odd.  Thus the same form as (\ref{simplifiedclassical}) is obtained, where only the $ x = 0 $ and $ x= 3 $ cases rotate the $ y $-ensemble:
\begin{align}
e^{-i H_{f_4} t}  | 0,1 \rangle \rangle | k_1 k_2 \rangle = 
\left\{
\begin{array}{cc}
| 0,1 \rangle \rangle | k_1 k_2 \rangle  & k_1,k_2 \leftrightarrow x= 1,2   \\
| 1,0 \rangle \rangle | k_1 k_2 \rangle  & k_1,k_2 \leftrightarrow x= 0,3 \\
\end{array}
\right. .
\end{align}

In ``quantum'' mode, after the Hadamard gates the initial state is (\ref{initialcondevenodd}), which in this case we write
\begin{align}
| \psi_{\text{init}} \rangle = | k_0 \rangle_x | \frac{1}{\sqrt{2}}, \frac{1}{\sqrt{2}} \rangle \rangle   | \frac{1}{\sqrt{2}}, \frac{1}{\sqrt{2}} \rangle \rangle  . 
\end{align}
Operating on this state, the state is 
\begin{align}
& e^{-i H_{f_4} t} | \psi_{\text{init}} \rangle  \nonumber \\
& =  \frac{1}{2} \exp \left[ -i \frac{\pi}{2} k_0 \left(  1 + (S^Z_1+N_1 +1)(S^Z_2+N_2 +1)  \right) \right] \nonumber \\
& \times | k_0 \rangle_x  \Big( | + \rangle \rangle  | + \rangle \rangle +
| + \rangle \rangle  | - \rangle \rangle + | - \rangle \rangle  | + \rangle \rangle
+ | - \rangle \rangle  |- \rangle \rangle \Big), 
\end{align}
where we have used the even and odd Schrodinger cat definitions of (\ref{evencat}) and (\ref{oddcat}).  
Since $ | + \rangle \rangle $ only contains even $ | k \rangle $ Fock states, and $ | - \rangle \rangle  $ contains odd $ | k \rangle $ Fock states, according to the same argument as (\ref{classicalexample}), the $ | + \rangle \rangle | + \rangle \rangle $ and $ | - \rangle \rangle | - \rangle \rangle $ terms pick up a factor of $ -1 $, while the other terms remain the same.  The state thus becomes
\begin{align}
 e^{-i H_{f_4} t} | \psi_{\text{init}} \rangle  &  =  - | k_0 \rangle_x \frac{1}{2} 
( | + \rangle \rangle  - | - \rangle \rangle )( | + \rangle \rangle  - | - \rangle \rangle ) \nonumber \\
& =  - | k_0 \rangle_x  | \frac{-1}{\sqrt{2}}, \frac{1}{\sqrt{2}} \rangle \rangle  | \frac{-1}{\sqrt{2}}, \frac{1}{\sqrt{2}} \rangle \rangle .
\label{ntwofinalstate}
\end{align}
The above state has zero overlap with the initial state $ | \psi_{\text{init}} \rangle $, which shows that in the ideal case this reproduces the Deutsch-Jozsa algorithm. 

Now let us introduce decoherence in the form of dephasing, which has a master equation \cite{byrnes2014,pyrkov14}
\begin{align}
\frac{d  \rho}{d t} = - \frac{\Gamma}{2} \sum_{n=1}^M \left( (S^Z_n )^2 \rho - 2 S^Z_n \rho S^Z_n + \rho (S^Z_n)^2 \right)
\end{align}
where $ \Gamma $ is the dephasing rate. For simplicity we ignore the dephasing on the $ y$-ensemble, as this takes a passive role operating in ``quantum'' mode.   The dephasing has the effect diminishing the off-diagonal terms
\begin{align}
\rho_{k_1 \dots k_M k_1' \dots k_M'}(t) = & \rho_{k_1 \dots k_M k_1' \dots  k_M'}(0)  e^{-2 \Gamma t \sum_{n=1}^M (k_n-k_n')^2} \nonumber  ,
\end{align}
where $ \rho (0) $ is the initial state and $ \rho_{k_1 \dots k_M k_1' \dots k_M'} = \langle k_1 \dots k_M | \rho | k_1' \dots k_M' \rangle $.  

In an experiment the observables are typically expectation values of the spin operators $ \langle S^{X,Y,Z} \rangle $.  As our aim is to distinguish between states where the state is preserved in $ | \psi_{\text{init}} \rangle $ (constant functions) and deviating from  $ | \psi_{\text{init}} \rangle $ (balanced functions), we define a signal quantity with respect to the initial state according to 
\begin{align}
{\cal S} = \prod_{n=1}^M \frac{1}{2} \left( 1 + \frac{\langle S^Z_n \rangle}{N_n} \right) .
\label{fidelityz}
\end{align}
where the expectation value is taken for the state at the end of the full gate sequence in Fig. \ref{fig1}(b).  
We can equally write the signal as 
\begin{align}
{\cal S} = \prod_{n=1}^M \frac{1}{2} \left( 1 + \frac{\langle S^X_n \rangle' }{N_n} \right) ,
\end{align}
where the state is taken to be immediately after the oracle.
For the constant cases, in the ideal case $ \langle S^Z_n \rangle/ N_n = 1 $ and we obtain $ {\cal S} = 1 $.  For the balanced cases, in the ideal case, in (\ref{afterhadamard}) all the terms give at least one spin where $ \langle S^Z_n \rangle/ N_n = -1 $, which immediately gives $ {\cal S} = 0 $.  This quantity may thus be used to distinguish between the constant and balanced cases. 

Let us examine what happens to the signal for each of the cases under the presence of decoherence, assumed to be present primarily during the oracle evaluation. In the constant case, $ | \psi_{\text{init}} \rangle $  remains unchanged due to the oracle.  Hence the only change that will occur to the $ x $-register in this case is the dephasing. The initial density matrix is thus
\begin{align}
\rho (0) = \prod_{n=1}^M |  \frac{1}{\sqrt{2}}, \frac{1}{\sqrt{2}} \rangle \rangle  \langle \langle  \frac{1}{\sqrt{2}}, \frac{1}{\sqrt{2}}| .
\end{align}
It is possible to evaluate exactly the time evolution under the master equation in this case, and we have 
\begin{align}
\langle S^X_n \rangle'   = N_n e^{-2 \Gamma t} .
\end{align}
The signal behaves as
\begin{align}
{\cal S}_{\text{constant}} = \left[ \frac{1}{2} ( 1+ e^{-2 \Gamma t} ) \right]^M \approx 1 - \Gamma M t ,
\label{fidelityconstant}
\end{align}
where we have assumed $ N_n = N $ for simplicity.  Note that there is no dependence on $ N $ for the signal, which shows that the same performance for macroscopic samples with large $ N $ are obtained as for qubits $ N = 1 $. The initial decay of the signal has a characteristic time $ t \sim 1/ \Gamma M $, which shows the signal is of the order of the dephasing time.  

For the balanced cases, we expect that the emergence of Schrodinger cat states will be very quickly destroyed into mixed states.  For example, for an initial state 
such as $ \frac{1}{\sqrt{2}} ( | 1,0 \rangle \rangle + | 0,1 \rangle \rangle ) $, the density matrix decays as 
\begin{align}
\rho(t) = & \frac{1}{2} \Big( | 1,0 \rangle \rangle \langle \langle 1,0 | + | 0,1 \rangle \rangle \langle \langle 0,1 | \nonumber \\
& +  e^{-2 N^2 \Gamma t} | 1,0 \rangle \rangle \langle \langle 0,1 | +  e^{-2 N^2 \Gamma t}  | 0,1 \rangle \rangle \langle \langle 1,0 | \Big)
\end{align}
which have off-diagonal terms that decay very quickly. This will be true for states such as (\ref{afterhadamard}), which is in general an entangled state involving Schrodinger cats.  In these cases, 
we would typically obtain a mixed state with expectations $ \langle S^Z_n \rangle/N_n \rightarrow 0 $.  Substituting into (\ref{fidelityz}), we expect the signal in these cases to be
\begin{align}
{\cal S}_{\text{balanced}} \approx \frac{1}{2^M} .
\label{fidelitybalanced}
\end{align}

Comparing (\ref{fidelityconstant}) and (\ref{fidelitybalanced}), we see that as long as  $ \Gamma M t \ll 1 $, it is possible to clearly distinguish between the constant and balanced cases, despite the presence of decoherence.  The reason for this is the fortuitous difference in the nature of the states in the balanced and constant cases.  In the constant cases, the states are untouched, hence the states remain spin coherent states, which are relatively stable states even in the presence of decoherence. Meanwhile, for the balanced cases, potentially decoherence-prone Schrodinger cat states are generated, which under decoherence evolve quickly to mixed states.  However, since the aim is to create a different state to the initial state, this mixed state is sufficient for detection of a balanced function. Thus while the decoherence indeed deteriorate the signal from the ideal value of $ {\cal S}=0 $ to $ {\cal S} = 1/ 2^M $, it does not do so in a catastrophic way.  Due to the nature of the detection of the Deutsch-Jozsa algorithm, despite the generation of fragile Schrodinger cat states, this allows for the detection to distinguish between the two cases.  Naturally this does not change the fact that Fock states need to be prepared for Method 1, which may be difficult in practice. Thus Method 2 may be the approach of choice for these considerations.

\section{Summary and Conclusions}
\label{sec:conc}

We have presented two methods of mapping the Deutsch-Jozsa algorithm, as originally formulated for qubits, onto implementations using ensembles of qubits.  We follow the EQC framework developed previously such that only Hamiltonians involving linear products of total spin operators are used, and collective measurements are made. In either of the two methods, the number of times the oracle needs to be executed is one, precisely the same as for the qubit case.  This provides an exponential quantum speedup over the classical case where at least half the input combinations must be tested.  The resource counts for the remaining part of the Deutsch-Jozsa algorithm is also the same, counting the resource for executing a Hadamard gate the same as for a qubit and ensemble.  

The two methods presented provide two different encodings for storing qubit information.  In Method 1, the binary information is stored as the parity of the Fock states.  The advantage of this approach is that it can map an arbitrary Deutsch-Jozsa oracle onto the EQC framework. The Deutsch-Jozsa oracle can be implemented using an infinite number of different Hamiltonians, and the approach is suitable if this generality is required in the mapping.  The drawback of Method 1 is that Schrodinger cat states are generated by the oracle, which are prone to decoherence.  However, as discussed in Sec. 
\ref{sec:decoherence}, due to the nature of the measurement discrimination between constant and balanced cases, in practice a clear signal difference should nevertheless be obtained between the two cases.  The reason for this is that to distinguish between the two cases, all that is required is a significant deviation from the initial state in the balanced case, which is realized even when decoherence is present.  

In Method 2, an encoding corresponding to orthogonal spin coherent states on the Bloch sphere was used.  This encoding cannot map all qubit oracle realizations, hence does not have the generality of Method 1.  It nonetheless can realize any of the $ {2^M \choose 2^{M-1} } $ balanced  and the two constant functions.  Hence the reduction in generality is only in the degrees of freedom allowed in the oracle realization, and not a restriction of the algorithm itself.  In this case, the algorithm works only to finite probability, hence is an approximation to the qubit case.  While approximate, the dominant errors are exponentially suppressed $ \sim e^{-0.77 N} $, hence in practice the errors are negligible for large ensemble sizes.  

This paper has shown that it is possible to perform the Deutsch-Jozsa using macroscopic ensembles under the practical restrictions imposed by EQC.  This joins the other quantum algorithms that are possible under EQC, namely quantum teleportation \cite{pyrkov14,pyrkov14b} and Deutsch's algorithm \cite{byrnes2014}.  Our results also reproduce the results already found for Deutsch's algorithm under a more general setting.  In the current work, our aim was simply to reproduce the results of the qubit version of the algorithm faithfully. One advantage of using ensembles is that it is possible -- unlike qubits -- to read out using non-destructive means the state of a spin coherent state \cite{ilookeke2014}.  Such features are not utilized in this or the other quantum algorithms that have been mapped successfully from qubits.  Utlizing such non-destructive measurements has the potential to lead to other quantum algorithms in EQC that are not possible with qubits.

\begin{acknowledgments}
This work is supported by the Shanghai Research Challenge Fund, New York University Global Seed Grants for Collaborative Research, National Natural Science Foundation of China grant 61571301, and the Thousand Talents Program for Distinguished Young Scholars. 
\end{acknowledgments}


\begin{thebibliography}{27}
\expandafter\ifx\csname natexlab\endcsname\relax\def\natexlab#1{#1}\fi
\expandafter\ifx\csname bibnamefont\endcsname\relax
  \def\bibnamefont#1{#1}\fi
\expandafter\ifx\csname bibfnamefont\endcsname\relax
  \def\bibfnamefont#1{#1}\fi
\expandafter\ifx\csname citenamefont\endcsname\relax
  \def\citenamefont#1{#1}\fi
\expandafter\ifx\csname url\endcsname\relax
  \def\url#1{\texttt{#1}}\fi
\expandafter\ifx\csname urlprefix\endcsname\relax\def\urlprefix{URL }\fi
\providecommand{\bibinfo}[2]{#2}
\providecommand{\eprint}[2][]{\url{#2}}

\bibitem[{\citenamefont{Deutsch}(1985)}]{deutsch1985}
\bibinfo{author}{\bibfnamefont{D.}~\bibnamefont{Deutsch}},
  \bibinfo{journal}{Proc. R. Soc. London A} \textbf{\bibinfo{volume}{400}},
  \bibinfo{pages}{97} (\bibinfo{year}{1985}).

\bibitem[{\citenamefont{Deutsch and Jozsa}(1992)}]{deutsch1992}
\bibinfo{author}{\bibfnamefont{D.}~\bibnamefont{Deutsch}} \bibnamefont{and}
  \bibinfo{author}{\bibfnamefont{R.}~\bibnamefont{Jozsa}},
  \bibinfo{journal}{Proc. R. Soc. London A} \textbf{\bibinfo{volume}{439}},
  \bibinfo{pages}{553} (\bibinfo{year}{1992}).

\bibitem[{\citenamefont{Cleve et~al.}(1998)\citenamefont{Cleve, Ekert,
  Macchiavello, and Mosca}}]{cleve1998}
\bibinfo{author}{\bibfnamefont{R.}~\bibnamefont{Cleve}},
  \bibinfo{author}{\bibfnamefont{A.}~\bibnamefont{Ekert}},
  \bibinfo{author}{\bibfnamefont{C.}~\bibnamefont{Macchiavello}},
  \bibnamefont{and} \bibinfo{author}{\bibfnamefont{M.}~\bibnamefont{Mosca}},
  \bibinfo{journal}{Proc. R. Soc. London A} \textbf{\bibinfo{volume}{454}},
  \bibinfo{pages}{339} (\bibinfo{year}{1998}).

\bibitem[{\citenamefont{Grover}(1996)}]{grover1996}
\bibinfo{author}{\bibfnamefont{L.~K.} \bibnamefont{Grover}}, in
  \emph{\bibinfo{booktitle}{Proceedings of the Twenty-eighth Annual ACM
  Symposium on Theory of Computing}} (\bibinfo{publisher}{ACM},
  \bibinfo{address}{New York, NY, USA}, \bibinfo{year}{1996}), STOC '96, pp.
  \bibinfo{pages}{212--219}, ISBN \bibinfo{isbn}{0-89791-785-5},
  \urlprefix\url{http://doi.acm.org/10.1145/237814.237866}.

\bibitem[{\citenamefont{Shor}(1994)}]{shor1994}
\bibinfo{author}{\bibfnamefont{P.~W.} \bibnamefont{Shor}}, in
  \emph{\bibinfo{booktitle}{Proceedings of the 35th Annual Symposium on
  Foundations of Computer Science}} (\bibinfo{publisher}{IEEE Computer
  Society}, \bibinfo{address}{Washington, DC, USA}, \bibinfo{year}{1994}), SFCS
  '94, pp. \bibinfo{pages}{124--134}, ISBN \bibinfo{isbn}{0-8186-6580-7},
  \urlprefix\url{http://dx.doi.org/10.1109/SFCS.1994.365700}.

\bibitem[{\citenamefont{Collins et~al.}(2000)\citenamefont{Collins, Kim,
  Holton, Sierzputowska-Gracz, and Stejskal}}]{collins2000}
\bibinfo{author}{\bibfnamefont{D.}~\bibnamefont{Collins}},
  \bibinfo{author}{\bibfnamefont{K.~W.} \bibnamefont{Kim}},
  \bibinfo{author}{\bibfnamefont{W.~C.} \bibnamefont{Holton}},
  \bibinfo{author}{\bibfnamefont{H.}~\bibnamefont{Sierzputowska-Gracz}},
  \bibnamefont{and} \bibinfo{author}{\bibfnamefont{E.~O.}
  \bibnamefont{Stejskal}}, \bibinfo{journal}{Phys. Rev. A}
  \textbf{\bibinfo{volume}{62}}, \bibinfo{pages}{022304}
  (\bibinfo{year}{2000}).

\bibitem[{\citenamefont{Wu et~al.}(2011)\citenamefont{Wu, Li, Zheng, Luo, Feng,
  and Peng}}]{wu2011}
\bibinfo{author}{\bibfnamefont{Z.}~\bibnamefont{Wu}},
  \bibinfo{author}{\bibfnamefont{J.}~\bibnamefont{Li}},
  \bibinfo{author}{\bibfnamefont{W.}~\bibnamefont{Zheng}},
  \bibinfo{author}{\bibfnamefont{J.}~\bibnamefont{Luo}},
  \bibinfo{author}{\bibfnamefont{M.}~\bibnamefont{Feng}}, \bibnamefont{and}
  \bibinfo{author}{\bibfnamefont{X.}~\bibnamefont{Peng}},
  \bibinfo{journal}{Phys. Rev. A} \textbf{\bibinfo{volume}{84}},
  \bibinfo{pages}{042312} (\bibinfo{year}{2011}).

\bibitem[{\citenamefont{Schuch and Siewert}(2002)}]{schuch2002}
\bibinfo{author}{\bibfnamefont{N.}~\bibnamefont{Schuch}} \bibnamefont{and}
  \bibinfo{author}{\bibfnamefont{J.}~\bibnamefont{Siewert}},
  \bibinfo{journal}{Phys. Stat. Sol. (b)} \textbf{\bibinfo{volume}{233}},
  \bibinfo{pages}{482–489} (\bibinfo{year}{2002}).

\bibitem[{\citenamefont{Takeuchi}(2000)}]{takeuchi2000}
\bibinfo{author}{\bibfnamefont{S.}~\bibnamefont{Takeuchi}},
  \bibinfo{journal}{Phys. Rev. A} \textbf{\bibinfo{volume}{62}},
  \bibinfo{pages}{032301} (\bibinfo{year}{2000}).

\bibitem[{\citenamefont{Gulde et~al.}(2003)\citenamefont{Gulde, Riebe,
  Lancaster, Becher, Eschner, H{\"a}ffner, Schmidt-Kaler, Chuang, and
  Blatt}}]{gulde2003}
\bibinfo{author}{\bibfnamefont{S.}~\bibnamefont{Gulde}},
  \bibinfo{author}{\bibfnamefont{M.}~\bibnamefont{Riebe}},
  \bibinfo{author}{\bibfnamefont{G.~P.~T.} \bibnamefont{Lancaster}},
  \bibinfo{author}{\bibfnamefont{C.}~\bibnamefont{Becher}},
  \bibinfo{author}{\bibfnamefont{J.}~\bibnamefont{Eschner}},
  \bibinfo{author}{\bibfnamefont{H.}~\bibnamefont{H{\"a}ffner}},
  \bibinfo{author}{\bibfnamefont{F.}~\bibnamefont{Schmidt-Kaler}},
  \bibinfo{author}{\bibfnamefont{I.~L.} \bibnamefont{Chuang}},
  \bibnamefont{and} \bibinfo{author}{\bibfnamefont{R.}~\bibnamefont{Blatt}},
  \bibinfo{journal}{Nature} \textbf{\bibinfo{volume}{421}}, \bibinfo{pages}{48}
  (\bibinfo{year}{2003}).

\bibitem[{\citenamefont{Braunstein and van Loock}(2005)}]{braunstein05}
\bibinfo{author}{\bibfnamefont{S.}~\bibnamefont{Braunstein}} \bibnamefont{and}
  \bibinfo{author}{\bibfnamefont{P.}~\bibnamefont{van Loock}},
  \bibinfo{journal}{Rev. Mod. Phys.} \textbf{\bibinfo{volume}{77}},
  \bibinfo{pages}{513} (\bibinfo{year}{2005}).

\bibitem[{\citenamefont{Julsgaard et~al.}(2001)\citenamefont{Julsgaard,
  Kozhekin, and Polzik}}]{julsgaard01}
\bibinfo{author}{\bibfnamefont{B.}~\bibnamefont{Julsgaard}},
  \bibinfo{author}{\bibfnamefont{A.}~\bibnamefont{Kozhekin}}, \bibnamefont{and}
  \bibinfo{author}{\bibfnamefont{E.~S.} \bibnamefont{Polzik}},
  \bibinfo{journal}{Nature} \textbf{\bibinfo{volume}{413}},
  \bibinfo{pages}{400} (\bibinfo{year}{2001}).

\bibitem[{\citenamefont{Krauter et~al.}(2012)\citenamefont{Krauter, Salart,
  Muschik, Petersen, Shen, Fernholz, and Polzik}}]{krauter12}
\bibinfo{author}{\bibfnamefont{H.}~\bibnamefont{Krauter}},
  \bibinfo{author}{\bibfnamefont{D.}~\bibnamefont{Salart}},
  \bibinfo{author}{\bibfnamefont{C.~A.} \bibnamefont{Muschik}},
  \bibinfo{author}{\bibfnamefont{J.~M.} \bibnamefont{Petersen}},
  \bibinfo{author}{\bibfnamefont{H.}~\bibnamefont{Shen}},
  \bibinfo{author}{\bibfnamefont{T.}~\bibnamefont{Fernholz}}, \bibnamefont{and}
  \bibinfo{author}{\bibfnamefont{E.~S.} \bibnamefont{Polzik}},
  \bibinfo{journal}{Nature Phys.} \textbf{\bibinfo{volume}{9}},
  \bibinfo{pages}{400} (\bibinfo{year}{2012}).

\bibitem[{\citenamefont{Byrnes et~al.}(2012)\citenamefont{Byrnes, Wen, and
  Yamamoto}}]{byrnes2012}
\bibinfo{author}{\bibfnamefont{T.}~\bibnamefont{Byrnes}},
  \bibinfo{author}{\bibfnamefont{K.}~\bibnamefont{Wen}}, \bibnamefont{and}
  \bibinfo{author}{\bibfnamefont{Y.}~\bibnamefont{Yamamoto}},
  \bibinfo{journal}{Phys. Rev. A} \textbf{\bibinfo{volume}{85}},
  \bibinfo{pages}{040306} (\bibinfo{year}{2012}).

\bibitem[{\citenamefont{Byrnes et~al.}(2015)\citenamefont{Byrnes, Rosseau,
  Khosla, Pyrkov, Thomasen, Mukai, Koyama, Abdelrahman, and
  Ilo-Okeke}}]{byrnes2014}
\bibinfo{author}{\bibfnamefont{T.}~\bibnamefont{Byrnes}},
  \bibinfo{author}{\bibfnamefont{D.}~\bibnamefont{Rosseau}},
  \bibinfo{author}{\bibfnamefont{M.}~\bibnamefont{Khosla}},
  \bibinfo{author}{\bibfnamefont{A.}~\bibnamefont{Pyrkov}},
  \bibinfo{author}{\bibfnamefont{A.}~\bibnamefont{Thomasen}},
  \bibinfo{author}{\bibfnamefont{T.}~\bibnamefont{Mukai}},
  \bibinfo{author}{\bibfnamefont{S.}~\bibnamefont{Koyama}},
  \bibinfo{author}{\bibfnamefont{A.}~\bibnamefont{Abdelrahman}},
  \bibnamefont{and}
  \bibinfo{author}{\bibfnamefont{E.}~\bibnamefont{Ilo-Okeke}},
  \bibinfo{journal}{Opt. Comm.} \textbf{\bibinfo{volume}{337}},
  \bibinfo{pages}{102} (\bibinfo{year}{2015}).

\bibitem[{\citenamefont{Pyrkov and Byrnes}(2014{\natexlab{a}})}]{pyrkov14}
\bibinfo{author}{\bibfnamefont{A.~N.} \bibnamefont{Pyrkov}} \bibnamefont{and}
  \bibinfo{author}{\bibfnamefont{T.}~\bibnamefont{Byrnes}},
  \bibinfo{journal}{New J. Phys.} \textbf{\bibinfo{volume}{16}},
  \bibinfo{pages}{073038} (\bibinfo{year}{2014}{\natexlab{a}}).

\bibitem[{\citenamefont{Pyrkov and Byrnes}(2014{\natexlab{b}})}]{pyrkov14b}
\bibinfo{author}{\bibfnamefont{A.~N.} \bibnamefont{Pyrkov}} \bibnamefont{and}
  \bibinfo{author}{\bibfnamefont{T.}~\bibnamefont{Byrnes}},
  \bibinfo{journal}{Phys. Rev. A.} \textbf{\bibinfo{volume}{90}},
  \bibinfo{pages}{062336} (\bibinfo{year}{2014}{\natexlab{b}}).

\bibitem[{\citenamefont{Bohi et~al.}(2009)\citenamefont{Bohi, Riedel,
  Hoffrogge, Reichel, Hansch, and Treutlein}}]{bohi09}
\bibinfo{author}{\bibfnamefont{P.}~\bibnamefont{Bohi}},
  \bibinfo{author}{\bibfnamefont{M.~F.} \bibnamefont{Riedel}},
  \bibinfo{author}{\bibfnamefont{J.}~\bibnamefont{Hoffrogge}},
  \bibinfo{author}{\bibfnamefont{J.}~\bibnamefont{Reichel}},
  \bibinfo{author}{\bibfnamefont{T.~W.} \bibnamefont{Hansch}},
  \bibnamefont{and}
  \bibinfo{author}{\bibfnamefont{P.}~\bibnamefont{Treutlein}},
  \bibinfo{journal}{Nature Phys.} \textbf{\bibinfo{volume}{5}},
  \bibinfo{pages}{592} (\bibinfo{year}{2009}).

\bibitem[{\citenamefont{Riedel et~al.}(2010)\citenamefont{Riedel, Bohi, Li,
  Hansch, Sinatra, and Treutlein}}]{riedel10}
\bibinfo{author}{\bibfnamefont{M.~F.} \bibnamefont{Riedel}},
  \bibinfo{author}{\bibfnamefont{P.}~\bibnamefont{Bohi}},
  \bibinfo{author}{\bibfnamefont{Y.}~\bibnamefont{Li}},
  \bibinfo{author}{\bibfnamefont{T.~W.} \bibnamefont{Hansch}},
  \bibinfo{author}{\bibfnamefont{A.}~\bibnamefont{Sinatra}}, \bibnamefont{and}
  \bibinfo{author}{\bibfnamefont{P.}~\bibnamefont{Treutlein}},
  \bibinfo{journal}{Nature} \textbf{\bibinfo{volume}{464}},
  \bibinfo{pages}{1170} (\bibinfo{year}{2010}).

\bibitem[{\citenamefont{Pyrkov and Byrnes}(2013)}]{pyrkov2013}
\bibinfo{author}{\bibfnamefont{A.}~\bibnamefont{Pyrkov}} \bibnamefont{and}
  \bibinfo{author}{\bibfnamefont{T.}~\bibnamefont{Byrnes}},
  \bibinfo{journal}{New J. Phys.} \textbf{\bibinfo{volume}{15}},
  \bibinfo{pages}{093019} (\bibinfo{year}{2013}).

\bibitem[{\citenamefont{Hussain et~al.}(2014)\citenamefont{Hussain, Ilo-Okeke,
  and Byrnes}}]{hussain2014}
\bibinfo{author}{\bibfnamefont{M.~I.} \bibnamefont{Hussain}},
  \bibinfo{author}{\bibfnamefont{E.~O.} \bibnamefont{Ilo-Okeke}},
  \bibnamefont{and} \bibinfo{author}{\bibfnamefont{T.}~\bibnamefont{Byrnes}},
  \bibinfo{journal}{Phys. Rev. A} \textbf{\bibinfo{volume}{89}},
  \bibinfo{pages}{053607} (\bibinfo{year}{2014}).

\bibitem[{\citenamefont{Abdelrahman et~al.}(2014)\citenamefont{Abdelrahman,
  Mukai, H{\"a}ffner, and Byrnes}}]{abdelrahman14}
\bibinfo{author}{\bibfnamefont{A.}~\bibnamefont{Abdelrahman}},
  \bibinfo{author}{\bibfnamefont{T.}~\bibnamefont{Mukai}},
  \bibinfo{author}{\bibfnamefont{H.}~\bibnamefont{H{\"a}ffner}},
  \bibnamefont{and} \bibinfo{author}{\bibfnamefont{T.}~\bibnamefont{Byrnes}},
  \bibinfo{journal}{Opt. Express} \textbf{\bibinfo{volume}{22}},
  \bibinfo{pages}{3501} (\bibinfo{year}{2014}).

\bibitem[{\citenamefont{Byrnes}(2013)}]{byrnes13}
\bibinfo{author}{\bibfnamefont{T.}~\bibnamefont{Byrnes}},
  \bibinfo{journal}{Phys. Rev. A} \textbf{\bibinfo{volume}{88}},
  \bibinfo{pages}{023609} (\bibinfo{year}{2013}).

\bibitem[{\citenamefont{Kurkjian et~al.}(2013)\citenamefont{Kurkjian,
  Pawłowski, Sinatra, and Treutlein}}]{kurkjian13}
\bibinfo{author}{\bibfnamefont{H.}~\bibnamefont{Kurkjian}},
  \bibinfo{author}{\bibfnamefont{K.}~\bibnamefont{Pawłowski}},
  \bibinfo{author}{\bibfnamefont{A.}~\bibnamefont{Sinatra}}, \bibnamefont{and}
  \bibinfo{author}{\bibfnamefont{P.}~\bibnamefont{Treutlein}},
  \bibinfo{journal}{Phys. Rev. A} \textbf{\bibinfo{volume}{88}},
  \bibinfo{pages}{043605} (\bibinfo{year}{2013}).

\bibitem[{\citenamefont{Byrnes et~al.}(2014)\citenamefont{Byrnes, Rosseau,
  Khosla, Pyrkov, Thomasen, Mukai, Koyama, Abdelrahman, and
  Ilo-Okeke}}]{byrnes14}
\bibinfo{author}{\bibfnamefont{T.}~\bibnamefont{Byrnes}},
  \bibinfo{author}{\bibfnamefont{D.}~\bibnamefont{Rosseau}},
  \bibinfo{author}{\bibfnamefont{M.}~\bibnamefont{Khosla}},
  \bibinfo{author}{\bibfnamefont{A.}~\bibnamefont{Pyrkov}},
  \bibinfo{author}{\bibfnamefont{A.}~\bibnamefont{Thomasen}},
  \bibinfo{author}{\bibfnamefont{T.}~\bibnamefont{Mukai}},
  \bibinfo{author}{\bibfnamefont{S.}~\bibnamefont{Koyama}},
  \bibinfo{author}{\bibfnamefont{A.}~\bibnamefont{Abdelrahman}},
  \bibnamefont{and}
  \bibinfo{author}{\bibfnamefont{E.}~\bibnamefont{Ilo-Okeke}},
  \bibinfo{journal}{Opt. Comm.} \textbf{\bibinfo{volume}{337}},
  \bibinfo{pages}{102} (\bibinfo{year}{2014}).

\bibitem[{\citenamefont{Ilo-Okeke and Byrnes}(2014)}]{ilookeke2014}
\bibinfo{author}{\bibfnamefont{E.~O.} \bibnamefont{Ilo-Okeke}}
  \bibnamefont{and} \bibinfo{author}{\bibfnamefont{T.}~\bibnamefont{Byrnes}},
  \bibinfo{journal}{Phys. Rev. Lett.} \textbf{\bibinfo{volume}{112}},
  \bibinfo{pages}{233602} (\bibinfo{year}{2014}).

\bibitem[{\citenamefont{Nielsen and Chuang}(2000)}]{nielsen00}
\bibinfo{author}{\bibfnamefont{M.~A.} \bibnamefont{Nielsen}} \bibnamefont{and}
  \bibinfo{author}{\bibfnamefont{I.~L.} \bibnamefont{Chuang}},
  \emph{\bibinfo{title}{Quantum Computation and Quantum Information,}}
  (\bibinfo{publisher}{Cambridge University Press}, \bibinfo{address}{New
  York}, \bibinfo{year}{2000}).

\end{thebibliography}

\end{document}